\documentclass[aps,prb,twocolumn,letterpaper,superscriptaddress]{revtex4-2}

\usepackage{amsmath}
\usepackage{amssymb}
\usepackage{amsfonts}
\usepackage{bm}
\usepackage{mathrsfs}
\usepackage{tensor}
\usepackage{dsfont}
\usepackage{esint}
\usepackage{comment}
\usepackage{graphicx}
\usepackage{xcolor}
\usepackage{dcolumn}

\usepackage[usenames,dvipsnames]{xcolor}
\usepackage[hyperindex,pdftex, breaklinks,colorlinks = true,linkcolor = blue,urlcolor=blue,citecolor=blue]{hyperref}

\usepackage{physics}
\usepackage[section]{placeins}

\begin{document}

	\title{Ferroelectric superconductivity in noncentrosymmetric metals}

	\author{Jason G. Kattan}
	\email{jason.kattan@mcgill.ca}
    \affiliation{Department of Physics, McGill University, Montréal, Québec H3A 2T8, Canada}

    \author{T. Pereg-Barnea}
    \email{tamar.pereg-barnea@mcgill.ca}
    \affiliation{Department of Physics, McGill University, Montréal, Québec H3A 2T8, Canada}
	
	\date{\today}
	
	\begin{abstract}
        It has recently been shown in experiments that certain materials can display both superconductivity and ferroelectricity, contrary to a long-standing conjecture that these two phenomena are incompatible, or at least unrelated. In this work we study superconductivity in ferroelectric metals, using a formalism of ionic polarization fields coupled to itinerant electrons, both of which are treated at the microscopic level. The ferroelectric order manifests as a spontaneous polarization that may be uniform or spatially modulated, and fluctuations of the polarization mediate interactions between the electrons. The polarization fluctuations give rise to attractive interactions that can lead to Cooper pairing in certain lattice configurations, analogous to the nonpolar phonons in conventional BCS theory. Working with a simplified BCS model, we derive conditions under which superconductivity can coexist with and even emerge from ferroelectricity.
    \end{abstract}
	
	\maketitle


\section{Introduction}\label{Sec:Introduction}

Ferroelectricity and superconductivity have long been viewed as incompatible phases of matter \cite{Anderson1965,Lines1977,Benedek2016,HickoxYoung2023}. The classical model of ferroelectricity in insulators involves a long-range alignment of electric dipoles that is stabilized by electrostatic interactions and manifests as a macroscopic spontaneous polarization, having the property that its direction can be switched by application of an electric field \cite{Resta1994,KingSmith1993}. Superconductivity arises from a metallic state with itinerant charge carriers that tend to rearrange themselves to screen the very electrostatic forces that would otherwise support a spontaneous polarization. Yet this expectation has become increasingly suspect; a growing body of experiments indicates that the proximity to a ferroelectric phase transition, and in some cases even a genuine switchable polarization, may survive in certain metals, and can even become correlated with the superconducting order
\cite{Rischau2017,Ahadi2019,Hameed2022,Fei2018,DeLaBarrera2021,Jindal2023,Zhang2026,Dong2026}.

Experiments on the oxide strontium titanate family provided some of the earliest indications that it may be possible for a ferroelectric material to become superconducting \cite{Schooley1964,Schooley1965}. The ground state of the quantum paraelectric SrTiO$_3$ lives near a ferroelectric instability \cite{Muller1979,Rowley2014}, and experiments on oxygen-deficient SrTiO$_3$ with calcium substitution revealed a ferroelectric quantum phase transition that occurs inside the superconducting dome \cite{Rischau2017}. There is even a finite region of the phase diagram in which dilute metallicity, superconductivity, and ferroelectricity may coexist \cite{Rischau2017}. Strain engineering enlarges this region: In strained SrTiO$_3$ thin films, the ferroelectric order is enhanced by an external straining force, and the critical temperature for the superconducting phase transition is found to increase substantially, reaching roughly twice the value of unstrained samples \cite{Ahadi2019,Russell2019}. This suggests that the low-energy polar degrees of freedom in these ``conventional ferroelectrics'' are not always mere spectators, and can even enhance superconductivity and reshape the phase diagram \cite{Dunnett2018,VanDerMarel2019,Gastiasoro2020,Saha2025}.

Another challenge to the classical expectation of competing orders comes from atomically thin van der Waals materials. For example, experiments on multilayer WTe$_2$ demonstrate a switchable out-of-plane polarization that persists despite the in-plane conducting properties, and subsequent capacitance measurements established bilayer WTe$_2$ as a tunable ferroelectric metal \cite{Fei2018,DeLaBarrera2021}. These studies demonstrate that metallic screening does not automatically destroy ferroelectricity when the material geometry, carrier density, and symmetry conditions are favorable, and the spontaneous polarization need not be generated by the usual ionic off-centering that is characteristic of conventional ferroelectrics. Instead, the broken inversion symmetry and switchable polarization arise from relative interlayer displacement, and an associated charge transfer between layers in a noncentrosymmetric stacking configuration
\cite{Yang2018,Liu2019,Wu2021,ViznerStern2021,Yasuda2021,Meng2022}. Certain members of this emerging class of ``sliding ferroelectrics'' have already been shown to host a superconducting phase; for example, simultaneous ferroelectric switching and superconductivity have been observed in bilayer $T_d$-MoTe$_2$, and a field-driven superconductor-to-normal transition was reported at the ferroelectric transition \cite{Jindal2023}. More recently, LaAlO$_3$/KTaO$_3$ interfaces have been shown to demonstrate a pronounced enhancement of the critical temperature \cite{Zhang2026,Kim2025}.

Existing phenomenological models of superconductivity in ferroelectrics, for example based on Ginzburg-Landau theory and its multiorder generalizations, are useful for building a macroscopic picture of what changes across the phase transition \cite{Ginzburg1950,Sigrist1991,Kanasugi2018}. But, by construction, they leave implicit any understanding of how the underlying ionic structure and Coulomb interactions conspire at the microscopic level to generate an attractive pairing interaction in the broken-symmetry phase. There have been a few recent theoretical studies that focus specifically on the pairing mechanism in sliding ferroelectrics, considering Cooper pair formation at domain walls between polarization domains, for example, or the role of interlayer charge-transfer mechanisms in this pairing \cite{Chaudhary2024,Biswas2026,Annenkov2025}. However, with the proliferation of experimental work searching for new materials that demonstrate this ``ferroelectric superconductivity," it would be useful to have a general formalism that is applicable to both conventional and sliding ferroelectrics, and in terms of which simple models could be constructed to establish conditions for the coexistence of ferroelectric and superconducting phases.

We introduce a Hamiltonian for ferroelectric superconductors that models both the electrons and ions at the microscopic level, where the degrees of freedom for spin-$1/2$ electrons are described by a second-quantized field operator, while the ionic degrees of freedom are encoded in a microscopic polarization field that is built from electric multipole moments for each of the ions in the crystal's lattice, instead of in a macroscopic coarse-grained way. This is motivated by the central role that polarization already plays in the theory of insulating ferroelectrics \cite{Lines1977,Resta1994,KingSmith1993}, but here it is elevated to a \textit{microscopic} field operator capable of describing both conventional and sliding ferroelectrics by an appropriate choice of spatial profile \cite{Liu2019,Wu2021,ViznerStern2021}; this kind of formalism based on microscopic notions of polarization and magnetization has been developed for atomic and molecular systems \cite{KattanSipe2023}, insulators \cite{Mahon2019}, metals \cite{Mahon2023,Kattan2026AHE}, and topological materials \cite{CImanuscript1,Kattan2}. The electrons and ions couple through the screened Coulomb interaction and spin-orbit coupling \cite{Gorkov2001,Frigeri2004,Kanasugi2018}.

We provide a field-theoretic description of the route from ferroelectricity to Cooper pairing in the following way. Oscillations of the ions about their polar equilibrium positions are described in terms of normal-mode fluctuations, which are phonons in the quantum theory. Integrating out these phonons leads to an attractive interaction between the electrons that competes with the repulsive Coulomb interaction. Just like in conventional Bardeen-Cooper-Schrieffer (BCS) theory, when the phonon-mediated interaction dominates over the Coulomb repulsion a Cooper instability may occur, leading to pairing between the electrons \cite{Frohlich1950,Bardeen1957,Morel1962}. Unlike in conventional BCS theory, however, our phonon-mediated interaction includes the ferroelectric order and general spin-orbit coupling \cite{Kanasugi2018,Saha2025}. Since the ionic degrees of freedom are described by a microscopic polarization field, our results can be naturally adapted to settings where the polar degree of freedom is geometric or interfacial, as in sliding ferroelectrics, as well as the usual ionic off-centering in conventional ferroelectrics.

When the net electronic interaction is attractive in the broken-symmetry phase, ferroelectric superconductivity may emerge \cite{Cooper1956,Bardeen1957,Tinkham1996}. Determining whether pairing will occur is an energy optimization problem.  To do so effectively, we need to move away from our very general interaction kernel and choose a more specific situation. We therefore reduce our model to only a single relevant band, neglect spin-orbit coupling, and project onto the Cooper channel in our interaction kernel. This allows us to find a polarization-dependent gap equation from which we can derive a set of inequalities that, when satisfied, describe the enhancement of superconductivity in a ferroelectric metal. We also provide a sample calculation of these inequalities in the simplified case of a circular Fermi surface, deriving a ferroelectric BCS theory \cite{Bardeen1957,Tinkham1996}. Further generalization is left for future work. 

We begin in Sec.~\ref{Sec:Fieldtheory} with a formulation of the second-quantized field theory in terms of which the coupled electron-ion dynamics is studied. Our Hamiltonian is very general, describing both the electrons and ions at the microscopic level, and includes both the screened Coulomb interactions and spin-orbit coupling in its general form. Next in Sec.~\ref{Sec:Ferroelectrics} we study the ferroelectric sector by introduction of a microscopic polarization field for the ions, with which we describe both the paraelectric and ferroelectric phases above and below the critical temperature for the phase transition. Then in Sec.~\ref{Sec:Pairing} we integrate out the phonons, resulting in an effective interaction between the electrons whose attractive eigenchannels may undergo the usual Cooper instability, which is reduced to a single-band ferroelectric BCS-type model. And this ferroelectric BCS model is used in Sec.~\ref{Sec:Practicalconsiderations} to formulate a set of inequalities for the enhancement of superconductivity in ferroelectric metals.

\section{Coupled electron-ion theory}\label{Sec:Fieldtheory}

To begin, we consider spin-$1/2$ electrons moving in a nominally infinite and regular lattice $\Gamma$ of ions that is extended in $d$ spatial dimensions. We treat the ionic degrees of freedom dynamically and subject to canonical quantization. Working in the Heisenberg picture, the electronic degrees of freedom are encoded in a pair of second-quantized field operators $\hat{\psi}_s(\bm{x},t)$ indexed by the spin index $s \in \{\uparrow,\downarrow\}$ and satisfying the equal-time anticommutation relations
\begin{align}
    \big\{\hat{\psi}_s(\bm{x},t), \hat{\psi}_{s'}^{\dagger}(\bm{y},t)\big\} &= \delta_{ss'} \delta(\bm{x} - \bm{y}),\nonumber \\
    \big\{\hat{\psi}_s(\bm{x},t), \hat{\psi}_{s'}(\bm{y},t)\big\} &= \big\{\hat{\psi}_s^{\dagger}(\bm{x},t), \hat{\psi}_{s'}^{\dagger}(\bm{y},t)\big\} = 0,
    \label{electronanticommutationcomponents}
\end{align}
which are collected into a $2$-spinor field operator
\begin{align}
    \hat{\psi}(\bm{x},t) \equiv \left(\mqty{\hat{\psi}_{\uparrow}(\bm{x},t) \\[2pt] \hat{\psi}_{\downarrow}(\bm{x},t)}\right).
\end{align}
This electron field operator encodes the dynamics of the valence and conduction electrons, whereas the strongly localized core electrons are modelled as being rigidly attached to the ions. Multilayer structures can be incorporated by supplementing the spin index with an additional layer (pseudospin) index \cite{Chaudhary2024}. 

The ionic degrees of freedom will be described in terms of first-quantized position and momentum operators, instead of field operators. Let $\bm{R} \in \Gamma$ denote the center of a given Wigner-Seitz cell, and let $M_I$ and $q_I$ denote the mass and charge of the $I\mathrm{th}$ ion in this unit cell, which moves about an equilibrium position $\bm{R}_I \equiv \bm{R} + \bm{d}_I$. To describe the ion dynamics about this equilibrium configuration, we define, for each ion, a \textit{deformation operator} $\hat{\bm{u}}_I(\bm{R},t)$ such that the position and momentum operators for this ion are
\begin{align}
    \hat{\bm{X}}_I(\bm{R},t) &= \bm{R}_I + \hat{\bm{u}}_I(\bm{R},t), \nonumber \\
    \hat{\bm{P}}_I(\bm{R},t) &= M_I \frac{\partial \hat{\bm{u}}_I(\bm{R},t)}{\partial t}.
    \label{ionposition}
\end{align}
These position and momentum operators satisfy the usual equal-time commutation relations for each ion separately; they commute between different ions, and they also commute with the electron field operator. 

To write down a Hamiltonian for this coupled electron-ion system, we introduce the electronic charge density
\begin{align}
    \hat{\rho}_e(\bm{x},t) = e \hat{\psi}^{\dagger}(\bm{x},t) \hat{\psi}(\bm{x},t),
    \label{electronicrho}
\end{align}
and, taking the ions to be pointlike, the ionic charge density is 
\begin{align}
    \hat{\rho}_{\mathrm{ion}}(\bm{x},t) = \sum_{I\bm{R}} q_I \delta(\bm{x} - \hat{\bm{X}}_I(\bm{R},t)).
    \label{ionicrho}
\end{align}
The total charge density is the sum
\begin{align}
    \hat{\rho}(\bm{x},t) = \hat{\rho}_e(\bm{x},t) + \hat{\rho}_{\mathrm{ion}}(\bm{x},t),
    \label{totalrho}
\end{align}
and the Hamiltonian is
\begin{align}
    \hat{H}(t) =&\; \frac{1}{2m} \int d\bm{x}\, \hat{\psi}^{\dagger}(\bm{x},t) \big(\bm{p}(\bm{x})\big)^2 \hat{\psi}(\bm{x},t)\nonumber \\
    &+ \sum_{I\bm{R}} \frac{1}{2M_I} \big(\hat{\bm{P}}_I(\bm{R},t)\big)^2 + \hat{U}(t),
    \label{Hinitial}
\end{align}
where $\bm{p}(\bm{x}) = (\hbar/i)\bm{\nabla}$ is the momentum operator for the electrons. The first two terms are the kinetic terms for the electrons and the ions, respectively, while the third term describes the interactions between them. The interaction term consists of two contributions, the first of which is the Coulomb interaction
\begin{align}
    \hat{U}_c(t) = \frac{1}{2} \int d\bm{x}d\bm{y}\, U(\bm{x} - \bm{y}) \hat{\rho}(\bm{x},t) \hat{\rho}(\bm{y},t),
\end{align}
involving the instantaneously screened Coulomb kernel
\begin{align}
    U(\bm{x} - \bm{y}) = \int \frac{d\bm{\kappa}}{(2\pi)^d} \frac{U_{\mathrm{bare}}(\bm{\kappa})}{1 - \Pi_{\mathrm{core}}(\bm{\kappa}) U_{\mathrm{bare}}(\bm{\kappa})} e^{i\bm{\kappa}\cdot(\bm{x} - \bm{y})},
    \label{Uscr}
\end{align}
where $U_{\mathrm{bare}}(\bm{\kappa})$ is the continuum Fourier transform of the bare Coulomb interaction, and $\Pi_{\mathrm{core}}(\bm{\kappa})$ is the static charge-density response function of the core electrons that are not described by the electron field operator. In terms of the electronic and ionic charge densities, this Coulomb interaction term is
\begin{align}
    \hat{U}_c(t) = \hat{U}_{e}(t) + \hat{U}_{\mathrm{ion}}(t) + \hat{U}_{e-\mathrm{ion}}(t),
\end{align}
where the electron-electron, ion-ion, and electron-ion interactions are given, respectively, by
\begin{align}
    &\hat{U}_{e}(t) = \frac{1}{2} \int d\bm{x}d\bm{y}\, U(\bm{x} - \bm{y})\hat{\rho}_e(\bm{x},t) \hat{\rho}_e(\bm{y},t),\label{Ue} \\
    &\hat{U}_{\mathrm{ion}}(t) = \frac{1}{2} \int d\bm{x}d\bm{y}\, U(\bm{x} - \bm{y})\hat{\rho}_{\mathrm{ion}}(\bm{x},t) \hat{\rho}_{\mathrm{ion}}(\bm{y},t),\label{Uion} \\
    &\hat{U}_{e-\mathrm{ion}}(t) = \int d\bm{x}d\bm{y}\, U(\bm{x} - \bm{y}) \hat{\rho}_{\mathrm{ion}}(\bm{x},t)\hat{\rho}_e(\bm{y},t).\label{Ueion}
\end{align}
The second contribution to the interaction term in the Hamiltonian (\ref{Hinitial}) is spin-orbit coupling, which we include to cover the possibility of unconventional pairing mechanisms \cite{Frigeri2004,Kanasugi2018}. For static ions, the standard spin-orbit coupling interaction is \cite{Kattan3}
\begin{align}
    \hat{U}_{\mathrm{soc}}(t) = \int d\bm{x}\, \hat{\psi}^{\dagger}(\bm{x},t) \mathcal{H}_{\mathrm{soc}}(\bm{x}) \hat{\psi}(\bm{x},t),
    \label{staticSOC}
\end{align}
where the spin-orbit coupling Hamiltonian density is
\begin{align}
    \mathcal{H}_{\mathrm{soc}}(\bm{x}) = \frac{\hbar}{4m^2 c^2} \bm{\sigma}\cdot\bm{\nabla}\mathrm{V}_{\Gamma}(\bm{x}) \times \bm{p}(\bm{x}),
\end{align}
involving the vector $\bm{\sigma} = (\sigma^x, \sigma^y, \sigma^z)$ of Pauli matrices, and $\mathrm{V}_{\Gamma}(\bm{x})$ is the potential energy generated by the static ions in the lattice. When considering dynamical ions, this potential energy becomes an operator
\begin{align}
    \hat{\mathrm{V}}(\bm{x},t) = e \int d\bm{y}\, U(\bm{x} - \bm{y}) \hat{\rho}_{\mathrm{ion}}(\bm{y},t),
\end{align}
and spin-orbit coupling becomes a genuine dynamical interaction between the electrons and the ions. Then the total interaction term is
\begin{align}
    \hat{U}(t) =&\, \frac{\hbar}{4m^2 c^2} \int d\bm{x}\, \hat{\psi}^{\dagger}(\bm{x},t) \Big(\bm{\sigma}\cdot\bm{\nabla}\hat{\mathrm{V}}(\bm{x},t) \times \bm{p}(\bm{x})\Big)\hat{\psi}(\bm{x},t)\nonumber \\
    &+ \int d\bm{x}\, \hat{\psi}^{\dagger}(\bm{x},t) \hat{\mathrm{V}}(\bm{x},t) \hat{\psi}(\bm{x},t) + \hat{U}_e(t) + \hat{U}_{\mathrm{ion}}(t),
    \label{totalinteraction}
\end{align}
which features in the Hamiltonian (\ref{Hinitial}).

\section{Ferroelectricity}\label{Sec:Ferroelectrics}

Ferroelectricity may arise when the equilibrium configuration of a crystal's lattice changes across a structural phase transition \cite{Cochran1960,Lines1977}. We take the paraelectric phase before the transition to be centrosymmetric, while in the ferroelectric phase after the transition there arises a polar distortion in the lattice that breaks inversion symmetry. This polar structure supports a spontaneous macroscopic polarization, and the phase is ferroelectric when this polarization can be switched between symmetry-related orientations by an electric field. And on either side of the phase transition, the ions undergo dynamical fluctuations about the corresponding equilibrium configuration.

The ion positions $\bm{R}_I = \bm{R} + \bm{d}_I$ introduced in Sec.~\ref{Sec:Fieldtheory} define a static configuration of the lattice for which the charge density is
\begin{align}
    \rho_{\mathrm{ion},0}(\bm{x}) = \sum_{I\bm{R}} q_I \delta(\bm{x} - \bm{R} - \bm{d}_I).
    \label{staticrhoion}
\end{align}
To account for the dynamical fluctuations around this static charge distribution, the ionic charge density operator is decomposed into this background density and a deformation-induced redistribution of charge, described by the fluctuation operator
\begin{align}
    \delta \hat{\rho}_{\mathrm{ion}}(\bm{x},t) = \hat{\rho}_{\mathrm{ion}}(\bm{x},t) - \rho_{\mathrm{ion},0}(\bm{x}).
\end{align}
Because a deformation in the lattice redistributes a fixed amount of charge throughout the crystal, the integral of this fluctuation operator vanishes, and therefore we can define a \textit{microscopic} polarization field operator for the ions relative to their equilibrium configuration through 
\begin{align}
    \delta \hat{\rho}_{\mathrm{ion}}(\bm{x},t) = - \bm{\nabla}\cdot\hat{\bm{p}}_{\mathrm{ion}}(\bm{x},t).
    \label{rhotop}
\end{align}
Notably, this polarization field is nonunique, and its explicit form can be chosen in such a way as to model both displacive and sliding ferroelectrics.

We write the ion terms in the Hamiltonian (\ref{Hinitial}) in terms of this polarization field. The kinetic energy of the ions is unchanged, while the ion-ion interaction term becomes
\begin{align}
    \hat{U}_{\mathrm{ion}}(t) =&\; E_{\mathrm{ion}} - \int d\bm{x}\, \hat{\bm{p}}_{\mathrm{ion}}(\bm{x},t)\cdot\bm{e}_{\mathrm{ion}}(\bm{x})\nonumber \\
    &+ \frac{1}{2} \int d\bm{x} d\bm{y}\, \hat{p}_{\mathrm{ion}}^i(\bm{x},t) H^{ij}(\bm{x} - \bm{y}) \hat{p}_{\mathrm{ion}}^j(\bm{y},t),
    \label{ionUp}
\end{align}
where the first term is a $c$-number energy associated with the static charge density (\ref{staticrhoion}), the second term couples the polarization field to the background electric field 
\begin{align}
    e_{\mathrm{ion}}^i(\bm{x}) = - \frac{\partial}{\partial x^i} \int d\bm{y}\, U(\bm{x} - \bm{y}) \rho_{\mathrm{ion},0}(\bm{y})
    \label{eion}
\end{align}
produced by this static charge density, and the third term involves the purely dynamical Coulomb interactions, involving the Hessian
\begin{align}
    H^{ij}(\bm{x} - \bm{y}) \equiv  \frac{\partial}{\partial x^i} \frac{\partial}{\partial y^j} U(\bm{x} - \bm{y})
    \label{HessianU}
\end{align}
of the screened Coulomb kernel (\ref{Uscr}). For the interaction term (\ref{totalinteraction}), we introduce the potential energy of the equilibrium lattice configuration,
\begin{align}
    \mathrm{V}_{\Gamma}(\bm{x}) = e \int d\bm{y}\, U(\bm{x} - \bm{y}) \rho_{\mathrm{ion},0}(\bm{y}),
\end{align}
and an effective electric field generated by the electrons,
\begin{align}
    \hat{e}^i(\bm{x},t) =&\; \frac{e \hbar\, \epsilon^{jk\ell} }{4m^2 c^2} \int d\bm{y}\, H^{ij}(\bm{x} - \bm{y})\hat{\psi}^{\dagger}(\bm{y},t) \sigma^{k} p^{\ell}(\bm{y}) \hat{\psi}(\bm{y},t)\nonumber \\
    &- \frac{\partial}{\partial x^i} \int d\bm{y}\, U(\bm{x} - \bm{y}) \hat{\rho}_e(\bm{y},t),
\end{align}
in terms of which the interaction term (\ref{totalinteraction}) becomes
\begin{align}
    \hat{U}(t) =&\; \frac{\hbar}{4m^2 c^2} \int d\bm{x}\, \hat{\psi}^{\dagger}(\bm{x},t) \Big(\bm{\sigma}\cdot\bm{\nabla}\mathrm{V}_{\Gamma}(\bm{x})\times\bm{p}(\bm{x})\Big)\hat{\psi}(\bm{x},t)\nonumber \\
    &+ \int d\bm{x}\, \hat{\psi}^{\dagger}(\bm{x},t) \mathrm{V}_{\Gamma}(\bm{x}) \hat{\psi}(\bm{x},t) + \hat{U}_e(t) + \hat{U}_{\mathrm{ion}}(t) \nonumber \\
    &- \int d\bm{x}\, \hat{\bm{p}}_{\mathrm{ion}}(\bm{x},t)\cdot\hat{\bm{e}}(\bm{x},t).
\end{align}
Then we can partition the Hamiltonian (\ref{Hinitial}) as
\begin{align}
    \hat{H}(t) = \hat{H}_e(t) + \hat{H}_{\mathrm{ion}}(t) + \hat{H}_{\mathrm{int}}(t),
    \label{Htot1}
\end{align}
where the electronic term is
\begin{align}
    \hat{H}_{e}(t) = \int d\bm{x}\, \hat{\psi}^{\dagger}(\bm{x},t) \mathcal{H}_e(\bm{x}) \hat{\psi}(\bm{x},t) + \hat{U}_e(t),
    \label{He}
\end{align}
involving the Hamiltonian density
\begin{align}
    \mathcal{H}_e(\bm{x}) =&\; \frac{1}{2m}\big(\bm{p}(\bm{x})\big)^2 + \mathrm{V}_{\Gamma}(\bm{x})\nonumber \\
    &+ \frac{\hbar}{4m^2 c^2} \bm{\sigma}\cdot\bm{\nabla}\mathrm{V}_{\Gamma}(\bm{x})\times\bm{p}(\bm{x}),
\end{align}
the ionic term is
\begin{align}
    \hat{H}_{\mathrm{ion}}(t) =&\; \sum_{I\bm{R}} \frac{1}{2M_I} \big(\hat{\bm{P}}_I(\bm{R},t)\big)^2 - \int d\bm{x}\, \hat{\bm{p}}_{\mathrm{ion}}(\bm{x},t)\cdot\bm{e}_{\mathrm{ion}}(\bm{x})\nonumber \\
    &+ \frac{1}{2} \int d\bm{x} d\bm{y}\, \hat{p}_{\mathrm{ion}}^i(\bm{x},t) H^{ij}(\bm{x} - \bm{y}) \hat{p}_{\mathrm{ion}}^j(\bm{y},t),
    \label{Hion}
\end{align}
where we have dropped the $c$-number energy $E_{\mathrm{ion}}$, and the interaction term takes the particularly simple form
\begin{align}
    \hat{H}_{\mathrm{int}}(t) = - \int d\bm{x}\, \hat{\bm{p}}_{\mathrm{ion}}(\bm{x},t)\cdot\hat{\bm{e}}(\bm{x},t).
    \label{eionUp}
\end{align}
We will analyze the paraelectric and ferroelectric phases in terms of this Hamiltonian.

\subsection{Paraelectric phase}

We first consider the paraelectric phase. The ion positions $\bm{R}_I = \bm{R} + \bm{d}_I$ are taken to correspond to an equilibrium configuration of the electron-ion system that is centrosymmetric, which fixes the expectation values
\begin{align}
    \langle\hat{\bm{u}}_I(\bm{R},t)\rangle_{\mathrm{PE}} &= \bm{0},\nonumber \\
    \langle\hat{\bm{X}}_I(\bm{R},t)\rangle_{\mathrm{PE}} &= \bm{R} + \bm{d}_I,
    \label{PEexpval}
\end{align}
for all ions $(I,\bm{R})$. Since these expectation values are taken in the paraelectric equilibrium state of the full coupled electron-ion system, the mean total force on each ion vanishes in this state; accordingly, the equilibrium expectation value of the Hamiltonian (\ref{Htot1}) contains no terms linear in any displacement about this configuration, although couplings linear in the deformation operators remain in the Hamiltonian itself. For small deformations of the ions about this equilibrium configuration, we expand the polarization field as
\begin{align}
    \hat{p}_{\mathrm{ion}}^i(\bm{x},t) = \sum_{n=0}^{\infty} \frac{1}{n!} \sum_{I\bm{R}}  \mathcal{D}^n p_{I}^{ij_1\dots j_n}(\bm{x},\bm{R}) \prod_{a=1}^n \hat{u}_{I}^{j_a}(\bm{R},t)
    \label{pexpansion}
\end{align}
where we have defined the functional derivatives
\begin{align}
    \mathcal{D}^{n} p_{I}^{i j_1\dots j_n}(\bm{x},\bm{R}) \equiv \eval{\frac{\delta^n p_{\mathrm{ion}}^i(\bm{x},t)}{\delta u_I^{j_1}(\bm{R},t) \dots \delta u_{I}^{j_n}(\bm{R},t)}}_{\bm{u}_I = \bm{0}},
\end{align}
which are evaluated at the expectation values (\ref{PEexpval}) in the paraelectric equilibrium state.

Considering the ion Hamiltonian (\ref{Hion}), we substitute this Taylor expansion into the second and third terms thereof, resulting in an expansion in powers of the deformation operators. It is useful to move to momentum space by implementing lattice Fourier decompositions
\begin{align}
    \hat{\bm{u}}_I(\bm{R},t) &= \frac{1}{\mathcal{N}_{\mathrm{uc}}} \int d\bm{q}\, e^{i\bm{q}\cdot\bm{R}} \hat{\bm{u}}_I(\bm{q},t),\nonumber \\
    \hat{\bm{P}}_I(\bm{R},t) &= \frac{1}{\mathcal{N}_{\mathrm{uc}}} \int d\bm{q}\, e^{i\bm{q}\cdot\bm{R}} \hat{\bm{P}}_I(\bm{q},t),
    \label{Fouriermodes}
\end{align}
where the normalization factor $\mathcal{N}_{\mathrm{uc}} = \sqrt{(2\pi)^d/\Omega_{\mathrm{uc}}}$ involves the volume $\Omega_{\mathrm{uc}}$ of the unit cell $\Omega$, and the integral is over the Brillouin zone $\mathrm{BZ}^d$. Dynamical fluctuations around the paraelectric background will be described in terms of normal modes, which are defined through the quadratic part of the ion Hamiltonian, 
\begin{align}
    \hat{H}_{\mathrm{ion}}^{(2)}(t) =&\; \sum_{I} \frac{1}{2M_I} \int d\bm{q}\, \hat{P}_{I}^i(\bm{q},t) \hat{P}_{I}^i(-\bm{q},t)\nonumber \\
    &+ \frac{1}{2} \sum_{IJ} \sqrt{M_I M_J} \int d\bm{q}\, F_{IJ}^{ij}(\bm{q}) \hat{u}_{I}^i(\bm{q},t) \hat{u}_{J}^j(-\bm{q},t),
\end{align}
where $F_{IJ}^{ij}(\bm{q})$ denotes the mass-weighted force constants matrix for fluctuations about the paraelectric equilibrium configuration (\ref{PEexpval}) of the full electron-ion system. It involves the polarization field and its derivatives, along with the static ionic background (\ref{eion}) and the electronic contribution to the equilibrium lattice energy. We emphasize that this is not a harmonic approximation to the full ion Hamiltonian; we diagonalize this force constants matrix, and write the deformation operators in the basis of eigenvectors thereof, both in the quadratic terms and the anharmonic terms in this Hamiltonian. We look for solutions of the eigenvalue equation
\begin{align}
    \sum_J F_{IJ}^{ij}(\bm{q}) e_{J\lambda}^j(\bm{q}) = \omega_{\lambda}(\bm{q})^2 e_{I\lambda}^i(\bm{q}),
\end{align}
subject to the orthonormality constraint
\begin{align}
    \sum_{I} e_{I\lambda}^i(\bm{q})^* e_{I\lambda'}^i(\bm{q}) = \delta_{\lambda\lambda'}.
\end{align}
If there are $N_{\mathrm{ion}}$ ions in the $d$-dimensional unit cell, then there are $dN_{\mathrm{ion}}$ basis vectors $\bm{e}_{I\lambda}(\bm{q})$, where the index $\lambda = 1,\dots, dN_{\mathrm{ion}}$ labels the corresponding normal-mode \textit{branches}. Write the Fourier modes (\ref{Fouriermodes}) as a linear combination of these basis vectors,
\begin{align}
    \hat{\bm{u}}_I(\bm{q},t) &= \frac{1}{\sqrt{M_I}} \sum_{\lambda} \bm{e}_{I\lambda}(\bm{q}) \hat{Q}_{\lambda}(\bm{q},t),\nonumber \\
    \hat{\bm{P}}_I(\bm{q},t) &= \sqrt{M_I} \sum_{\lambda} \bm{e}_{I\lambda}(\bm{q}) \hat{\Pi}_{\lambda}(\bm{q},t),
    \label{Fouriertonormal}
\end{align}
where the expansion ``coefficients" are the normal-mode operators
\begin{align}
    \hat{Q}_{\lambda}(\bm{q},t) &= \frac{1}{\mathcal{N}_{\mathrm{uc}}} \sum_{I\bm{R}} \sqrt{M_I} e^{-i\bm{q}\cdot\bm{R}} \bm{e}_{I\lambda}^*(\bm{q})\cdot\hat{\bm{u}}_I(\bm{R},t),\nonumber \\
    \hat{\Pi}_{\lambda}(\bm{q},t) &= \frac{1}{\mathcal{N}_{\mathrm{uc}}} \sum_{I\bm{R}} \frac{1}{\sqrt{M_I}} e^{-i\bm{q}\cdot\bm{R}} \bm{e}_{I\lambda}^*(\bm{q}) \cdot \hat{\bm{P}}_I(\bm{R},t).
    \label{normalmodeops}
\end{align}
Written in terms of these normal modes, the quadratic part of the ion Hamiltonian is
\begin{align}
    \hat{H}_{\mathrm{ion}}^{(2)}(t) =&\; \frac{1}{2} \sum_{\lambda} \int d\bm{q}\, \hat{\Pi}_{\lambda}(\bm{q},t) \hat{\Pi}_{\lambda}(-\bm{q},t)\nonumber \\
    &+ \frac{1}{2} \sum_{\lambda} \int d\bm{q}\, \omega_{\lambda}(\bm{q})^2 \hat{Q}_{\lambda}(\bm{q},t) \hat{Q}_{\lambda}(-\bm{q},t),
\end{align}
and, together with the anharmonic terms that result from higher-order powers in the expansion (\ref{pexpansion}), the total ion Hamiltonian (\ref{Hion}) becomes
\begin{align}
    \hat{H}_{\mathrm{ion}}(t) = \frac{1}{2} \sum_{\lambda} \int d\bm{q}\, \hat{\Pi}_{\lambda}(\bm{q},t) \hat{\Pi}_{\lambda}(-\bm{q},t) + \mathrm{V}_{\mathrm{ion}}(\hat{Q}),
    \label{Hion2}
\end{align}
where the ion potential energy is
\begin{widetext}
\begin{align}
    \mathrm{V}_{\mathrm{ion}}(\hat{Q}) = \frac{1}{2} \sum_{\lambda} \int d\bm{q}\, \omega_{\lambda}(\bm{q})^2 \hat{Q}_{\lambda}(\bm{q},t) \hat{Q}_{\lambda}(-\bm{q},t) + \sum_{n=3}^{\infty} \frac{1}{n!}\sum_{\{\lambda\}}  \int d\bm{q}_1 \dots d\bm{q}_n\, \Gamma_{n}^{\lambda_1\dots\lambda_n}(\bm{q}_1,\dots,\bm{q}_n)\prod_{a=1}^n \hat{Q}_{\lambda_a}(\bm{q}_a,t).
    \label{Viongeneric}
\end{align}
\\
\end{widetext}
The coefficients of the anharmonic terms can be calculated in a straightforward way by substituting the polarization expansion (\ref{pexpansion}) into the ion Hamiltonian (\ref{Hion}), implementing the Fourier decompositions (\ref{Fouriermodes}), and writing each term in the normal-mode basis using Eq.~(\ref{Fouriertonormal}). Lattice translation symmetry is contained in these coefficients: They vanish unless the normal-mode momenta sum to a reciprocal lattice vector, so crystal momentum is conserved modulo a reciprocal lattice vector.

We retain all normal-mode branches as dynamical lattice modes, but we work with a simplified even-order anharmonic model in which odd-order couplings between the normal modes are neglected. Although inversion symmetry does not necessarily eliminate every odd-order multimode coupling in a generic crystal, this approximation is appropriate when the lattice distortions and parity-mixing effects produced by these couplings are weak. Accordingly, in this approximation we set
\begin{align}
    \Gamma_{n}^{\lambda_1\dots\lambda_n}(\bm{q}_1, \dots, \bm{q}_n) = 0 \;\;\; \text{for all odd $n$},
\end{align}
and truncate the potential at quartic order, writing
\begin{widetext}
\begin{align}
    \mathrm{V}_{\mathrm{ion}}(\hat{Q}) = \frac{1}{2} \sum_{\lambda} \int d\bm{q}\, \omega_{\lambda}(\bm{q})^2 \hat{Q}_{\lambda}(\bm{q},t)\hat{Q}_{\lambda}(-\bm{q},t) + \frac{1}{4!} \sum_{\{\lambda\}} \int d\bm{q}_1 \dots d\bm{q}_4\, \Gamma_4^{\lambda_1\lambda_2\lambda_3\lambda_4}(\bm{q}_1,\bm{q}_2,\bm{q}_3,\bm{q}_4) \prod_{a = 1}^{4} \hat{Q}_{\lambda_a}(\bm{q}_a,t),
    \label{Vinversion}
\end{align}
\end{widetext}
which is sufficient for describing spontaneous symmetry breaking \cite{Lines1977}. The expectation values of the normal-mode operators in the paraelectric phase vanish,
\begin{align}
    \langle\hat{Q}_{\lambda}(\bm{q},t)\rangle_{\mathrm{PE}} = 0
\end{align}
for all normal modes $(\lambda,\bm{q})$. The ferroelectric phase is distinguished by the appearance of a nonzero expectation value for one or more of these normal modes.

\subsection{Ferroelectric phase}\label{Sec:Paraelectricferroelectricphases}

In the ferroelectric phase, a macroscopic polarization arises from a structural deformation in the crystal's lattice that breaks inversion symmetry. At the microscopic level, this spontaneous polarization is a consequence of a change in the paraelectric expectation values (\ref{PEexpval}). Assuming the macroscopic polarization is \textit{uniform}, the ferroelectric expectation values are
\begin{align}
    \langle\hat{\bm{u}}_I(\bm{R},t)\rangle_{\mathrm{FE}} &= \bm{u}_I,\nonumber \\
    \langle\hat{\bm{X}}_I(\bm{R},t)\rangle_{\mathrm{FE}} &= \bm{R} + \bm{d}_I + \bm{u}_I,
    \label{FEexpvals}
\end{align}
where $\bm{u}_I$ is the displacement of the $I\mathrm{th}$ ion in the unit cell from its paraelectric equilibrium position. Different ions within the unit cell may acquire different displacement vectors, so the microscopic polarization field generally retains spatial structure within each unit cell. However, these expectation values correspond to a uniform \textit{macroscopic} polarization because the displacement $\bm{u}_I$ does not change from one unit cell to another.

We express this polar distortion in the normal-mode basis of the paraelectric phase. Taking the ferroelectric expectation value of the normal-mode operators (\ref{normalmodeops}) and using that $\bm{u}_I$ is independent of $\bm{R}$ gives
\begin{align}
    \langle\hat{Q}_{\lambda}(\bm{q},t)\rangle_{\mathrm{FE}} = Q_{\lambda} \delta(\bm{q}),
\end{align}
where for each branch $\lambda$ we have defined
\begin{align}
    Q_{\lambda} \equiv \mathcal{N}_{\mathrm{uc}} \sum_I \sqrt{M_I}\bm{e}_{I\lambda}^*(\bm{0})\cdot\bm{u}_I.
\end{align}
Thus, the uniformity of the spontaneous polarization implies that one or more zone-center branches acquire a nonzero equilibrium expectation value. To simplify our model, we assume that this distortion is carried by a single branch $\lambda = \lambda_0$ that becomes polar. Then the normal-mode expectation values in the ferroelectric phase are
\begin{align}
    \langle\hat{Q}_{\lambda}(\bm{q},t)\rangle_{\mathrm{FE}} = Q_0 \delta(\bm{q}) \delta_{\lambda\lambda_{0}},
\end{align}
involving the \textit{ferroelectric order parameter}
\begin{align}
    Q_{0} \equiv Q_{\lambda_0} = \mathcal{N}_{\mathrm{uc}} \sum_I \sqrt{M_I}\bm{e}_{I\lambda_0}^*(\bm{0})\cdot\bm{u}_I.
    \label{Q0orderparameter}
\end{align}
This is a signed order parameter that vanishes in the paraelectric phase; the application of an electric field can switch between $\pm Q_0$ by changing the sign of the displacement vectors \cite{Fei2018,Jindal2023}. 

We expand the potential energy (\ref{Vinversion}) around the new stable equilibrium configuration in the ferroelectric phase, writing the normal-mode operators as
\begin{align}
    \hat{Q}_{\lambda}(\bm{q},t) &= \langle\hat{Q}_{\lambda}(\bm{q},t)\rangle_{\mathrm{FE}} + \delta\hat{Q}_{\lambda}(\bm{q},t)\nonumber \\
    &= Q_0 \delta(\bm{q}) \delta_{\lambda\lambda_{0}} + \delta\hat{Q}_{\lambda}(\bm{q},t).
    \label{deltaQdef}
\end{align}
The operators $\delta\hat{Q}_{\lambda}(\bm{q},t)$ and $\delta\hat{\Pi}_{\lambda}(\bm{q},t) = \hat{\Pi}_{\lambda}(\bm{q},t)$ describe fluctuations of the ions about their new polar lattice positions. This being a ``stable" equilibrium configuration means that the mean total force on each ion must vanish in this phase, and, accordingly, there are no terms linear in the ion displacements in the ferroelectric expectation value of the ion potential (\ref{Vinversion}). To further simplify our model, we neglect any quadratic mixing between the different normal-mode branches, and we also neglect the remaining cubic and quartic terms in the fluctuation operators above. The ion potential then becomes
\begin{align}
    \mathrm{V}_{\mathrm{ion}}(\hat{Q}) =&\; \frac{1}{2} \sum_{\lambda} \int d\bm{q}\, \Omega_{\lambda}(\bm{q})^2 \delta\hat{Q}_{\lambda}(\bm{q},t) \delta\hat{Q}_{\lambda}(-\bm{q},t),
    \label{ionicHfluct}
\end{align}
involving the renormalized frequencies
\begin{align}
    \Omega_{\lambda}(\bm{q})^2 = \omega_{\lambda}(\bm{q})^2 + \frac{1}{2} \Gamma_{4,\mathrm{sym}}^{\lambda_0\lambda_0\lambda\lambda}(\bm{0},\bm{0},\bm{q},-\bm{q}) Q_0^2,
    \label{shiftedfrequencies}
\end{align}
where the subscript ``sym" indicates that the quartic coefficient in the second term is completely symmetrized with respect to its arguments. 

We perform the same analysis for the electron-ion interaction term, writing it in terms of the normal-mode operators (\ref{normalmodeops}), and expanding around the ferroelectric background using Eq. (\ref{deltaQdef}). The ``renormalized" Hamiltonian for the ferroelectric phase that follows is
\begin{align}
    \hat{H}(t) = \hat{H}_e^R(t) + \hat{H}_{\mathrm{ion}}^R(t) + \hat{H}_{\mathrm{int}}^R(t),
    \label{Hferro}
\end{align}
where the electronic term is
\begin{align}
    \hat{H}_{e}^R(t) = \int d\bm{x}\, \hat{\psi}^{\dagger}(\bm{x},t) \mathcal{H}_e^R(\bm{x}) \hat{\psi}(\bm{x},t) + \hat{U}_e(t),
    \label{electronH}
\end{align}
involving a new Hamiltonian density
\begin{align}
    \mathcal{H}_e^R(\bm{x}) =&\; \frac{1}{2m}\big(\bm{p}(\bm{x})\big)^2 + \mathrm{V}_{\Gamma}^R(\bm{x})\nonumber \\
    &+ \frac{\hbar}{4m^2 c^2}  \bm{\sigma} \cdot \bm{\nabla}\mathrm{V}_{\Gamma}^R(\bm{x}) \times \bm{p}(\bm{x})
    \label{H0density}
\end{align}
that features the renormalized lattice potential energy
\begin{align}
    \mathrm{V}_{\Gamma}^R(\bm{x}) \equiv e\sum_{I\bm{R}} q_I U(\bm{x} - \bm{R}_I - \bm{u}_I).
    \label{Vrenorm}
\end{align}
The second term in the Hamiltonian (\ref{Hferro}) describes the fluctuations of the ions around the ferroelectric equilibrium configuration, given by
\begin{align}
    \hat{H}_{\mathrm{ion}}^R(t) =&\; \frac{1}{2} \sum_{\lambda} \int d\bm{q}\, \delta \hat{\Pi}_{\lambda}(\bm{q},t) \delta \hat{\Pi}_{\lambda}(-\bm{q},t)\nonumber \\
    &+ \frac{1}{2} \sum_{\lambda} \int d\bm{q}\, \Omega_{\lambda}(\bm{q})^2 \delta\hat{Q}_{\lambda}(\bm{q},t) \delta\hat{Q}_{\lambda}(-\bm{q},t).
    \label{ionHnondiag}
\end{align}
And the third term describes interactions between the electrons and the ions in the ferroelectric phase; truncating at linear order in the fluctuation operators,
\begin{align}
    \hat{H}_{\mathrm{int}}^R(t) = - \sum_{\lambda} \int d\bm{x} d\bm{q}\, \hat{\psi}^{\dagger}(\bm{x},t) \mathcal{G}_{\lambda}(\bm{x},\bm{q}) \hat{\psi}(\bm{x},t) \delta\hat{Q}_{\lambda}(\bm{q},t),
    \label{Hint}
\end{align}
where the linear electron-ion coupling is
\begin{align}
    \mathcal{G}_{\lambda}(\bm{x},\bm{q}) = \mathcal{G}_{\lambda}^{c}(\bm{x},\bm{q}) + \mathcal{G}_{\lambda}^{\nabla}(\bm{x},\bm{q}),
    \label{mathcalGtot}
\end{align}
involving the Coulomb and spin-orbit couplings
\begin{align}
    \mathcal{G}_{\lambda}^c(\bm{x},\bm{q}) &= \frac{e}{\mathcal{N}_{\mathrm{uc}}} \sum_{I\bm{R}} \frac{q_I}{\sqrt{M_I}} e^{i\bm{q}\cdot\bm{R}} \,\bm{e}_{I\lambda}(\bm{q}) \cdot \bm{\nabla} U(\bm{x} - \bm{R}_I - \bm{u}_I),\nonumber \\
    \mathcal{G}_{\lambda}^{\nabla}(\bm{x},\bm{q}) &= \frac{\hbar}{4m^2 c^2} \bm{\sigma} \cdot \bm{\nabla} \mathcal{G}_{\lambda}^c(\bm{x},\bm{q}) \times \bm{p}(\bm{x}).
    \label{Gtensorcontributions}
\end{align}
These kernels can also be written in terms of the ion polarization field, but this form will be more useful below.

\section{Superconductivity}\label{Sec:Pairing}

The most commonly observed mechanism that leads to superconductivity is the phonon-mediated Cooper instability, in which the screened Coulomb interactions that repel nearby electrons in a metal are overcome by attractive ion-mediated interactions between the electrons in at least one pairing channel
\cite{Bardeen1957,Tinkham1996}. In the quantum theory, normal-mode oscillations of the ions are phonons, and the attractive interactions mediated by these phonons may be enhanced or suppressed when a material undergoes a structural phase transition and enters a ferroelectric phase. The idea will be to ``integrate out" the phonons in the ferroelectric Hamiltonian (\ref{Hferro}), leading to an effective $2$-body interaction term for the electrons. Cooper pairing is favoured when the 
effective interaction develops an attractive eigenchannel, corresponding to a negative eigenvalue of the interaction kernel, resulting in the phonon-mediated Cooper instability that is associated with a superconducting phase transition.

The details of this calculation can be found in Appendix \ref{Sec:EffectiveHamiltonian}. There we introduce phonon creation and annihilation operators that are defined in terms of the normal-mode operators and their conjugate momenta, and, by solving the Heisenberg equation for these operators and substituting their solutions back into the Hamiltonian, we identify an effective $2$-body interaction term that describes a competition between the attractive phonon-mediated interactions and the repulsive screened Coulomb interactions. This very general effective electronic Hamiltonian written in a basis-independent way in terms of the spinor electron field operators is given in Appendix \ref{Sec:EffectiveHamiltonian}. However, it is useful to expand the field operators in terms of the Bloch energy eigenfunctions
\begin{align}
    \psi_{n\bm{k}}(\bm{x}) = \frac{1}{(2\pi)^{d/2}} e^{i\bm{k}\cdot\bm{x}} u_{n\bm{k}}(\bm{x})
    \label{Blocheigenfunctions}
\end{align}
where the cell-periodic $2$-spinor Bloch functions are
\begin{align}
    u_{n\bm{k}}(\bm{x}) = \left(\mqty{u_{n\bm{k}}^{\uparrow}(\bm{x}) \\[1pt] u_{n\bm{k}}^{\downarrow}(\bm{x})}\right),
\end{align}
in terms of which this effective electronic Hamiltonian is
\begin{widetext}
\begin{align}
    \hat{H}_{\mathrm{eff}}(t) = \sum_{n} \int d\bm{k}\, \varepsilon_{n}(\bm{k}) \hat{c}_{n\bm{k}}^{\dagger}(t) \hat{c}_{n\bm{k}}(t) + \frac{1}{2} \sum_{\{n\}} \int  \frac{d\bm{q}}{(2\pi)^d} d\bm{k} d\bm{k}'\, \mathcal{V}_{n_1 n_2 n_3 n_4}(\bm{k},\bm{k}',\bm{q}) \hat{c}_{n_1,\bm{k}+\bm{q}}^{\dagger}(t) \hat{c}_{n_2,\bm{k}'-\bm{q}}^{\dagger}(t) \hat{c}_{n_3\bm{k}'}(t) \hat{c}_{n_4\bm{k}}(t).
    \label{Hintkspace}
\end{align}
The electronic creation and annihilation operators satisfy the canonical anticommutation relations, and $\varepsilon_{n}(\bm{k})$ are the band energies. The second term features the $2$-body interaction kernel, which in the static limit is given by
\begin{align}
    \mathcal{V}_{n_1 n_2 n_3 n_4}(\bm{k},\bm{k}',\bm{q}) = \mathcal{V}_{n_1 n_2 n_3 n_4}^{(c)}(\bm{k},\bm{k}',\bm{q}) - \sum_{\lambda} \frac{1}{\Omega_{\lambda}(\bm{q})^2} G_{\lambda,n_1 n_4}(\bm{k},\bm{q}) G_{\lambda,n_2 n_3}(\bm{k}',-\bm{q}),
    \label{Vintkernel}
\end{align}
where the first term is the multiband Coulomb contribution
\begin{align}
    \mathcal{V}_{n_1 n_2 n_3 n_4}^{(c)}(\bm{k},\bm{k}',\bm{q}) = \frac{e^2}{\Omega_{\mathrm{uc}}} \sum_{\bm{R}} \int_{\Omega} d\bm{x} d\bm{y}\, e^{-i\bm{q}\cdot(\bm{x} -\bm{y} + \bm{R})} U(\bm{x} - \bm{y} + \bm{R}) u_{n_1,\bm{k} + \bm{q}}^{\dagger}(\bm{x}) u_{n_2,\bm{k}'-\bm{q}}^{\dagger}(\bm{y}) u_{n_3 \bm{k}' }(\bm{y}) u_{n_4 \bm{k}}(\bm{x}),
    \label{CoulombV}
\end{align}
\end{widetext}
while the second term involves the renormalized frequencies (\ref{shiftedfrequencies}) and the electron-phonon vertex factors
\begin{align}
    G_{\lambda,n_1 n_2}(\bm{k},\bm{q}) =  \int_{\Omega} d\bm{x}\, e^{-i\bm{q}\cdot\bm{x}} u_{n_1,\bm{k}+\bm{q}}^{\dagger}(\bm{x}) \mathcal{G}_{\lambda}(\bm{x},\bm{q}) u_{n_2\bm{k}}(\bm{x}).
    \label{electronphononvertex}
\end{align}
This couples an electron of momentum $\bm{k}$ to an electron of momentum $\bm{k} + \bm{q}$ through the background phonons.

This effective Hamiltonian can be used to study electron dynamics and superconductivity in a generic multiband system, for example, using \textit{ab initio} methods such as density functional theory \cite{Baroni2001,Giustino2017}. It describes the physics of electrons moving in a ferroelectric lattice that is dynamical, with collective lattice excitations that generate phonon-mediated interactions between the electrons. No assumption has been made about the spin structure, band content, or orbital symmetries of the electronic wavefunctions; the only assumption here is that the attractive interaction responsible for Cooper pairing arises from virtual excitations in the crystal's lattice. It is not restricted to the spin-independent superconductivity of conventional BCS theory; instead, it produces a very general phonon-mediated, spin-dependent, and possibly multiband interaction kernel whose attractive eigenchannel(s) may correspond to singlet, triplet, mixed-parity, or even spin-orbit-entangled pairing states.

This level of generality is useful for establishing the microscopic origin of the phonon-mediated Cooper instability, but it is far too broad to yield simple analytic conditions for when ferroelectricity and superconductivity may coexist. We therefore reduce this Hamiltonian down to a {\it ferroelectric BCS model} that makes transparent how the polar distortion modifies the band dispersion, phonon frequencies, and phonon-mediated interactions between electrons in a conventional BCS superconductor \cite{Tinkham1996,Kanasugi2018}. More precisely, this should be understood as a parametrized family of BCS models depending continuously on a real-valued variable $Q$ that parametrizes a family of uniform ferroelectric backgrounds. For each value of this parameter $Q$ the polar distortion defines a different BCS instability problem, but for a given material only $Q = Q_0$ is required to be its physical equilibrium order parameter. 

We begin with three simplifications. First, we restrict to a single electronic band $n = n_0$ that crosses the Fermi level and is assumed to dominate the Cooper instability. We also project the interaction kernel (\ref{Vintkernel}) for this single-band model onto the Cooper channel, where a Cooper pair with momenta $\bm{k}$ and $-\bm{k}$ scatters into a Cooper pair with momenta $\bm{p}$ and $-\bm{p}$ so that the center-of-mass momentum of the pair vanishes. And we neglect spin-orbit coupling in the electron-phonon couplings (\ref{Gtensorcontributions}), restricting to spin-independent interactions between the electrons. Under these simplifications, the parametrized interaction kernel becomes 
\begin{align}
     \mathcal{V}(\bm{k},\bm{p};Q) \equiv&\; \mathcal{V}^{(c)}(\bm{k},-\bm{k},\bm{q};Q)\nonumber \\
     &- \sum_{\lambda} \frac{1}{\Omega_{\lambda}(\bm{q};Q)^2} G_{\lambda}(\bm{k},\bm{q};Q) G_{\lambda}(-\bm{k},-\bm{q};Q),
     \label{BCSinteraction}
\end{align}
where $\bm{q} = \bm{p} - \bm{k}$ is the momentum transferred between the ingoing and outgoing Cooper pairs. 

We work in the Schrödinger picture with the grand-canonical Hamiltonian $\hat{K}_{\mathrm{eff}}(Q) = \hat{H}_{\mathrm{eff}}(Q) - \mu(Q) \hat{N}$, where $\hat{N}$ is the total particle number operator, and $\mu(Q)$ is the background-dependent chemical potential. Following the standard mean-field decoupling \cite{Bardeen1957,Tinkham1996}, this Hamiltonian becomes
\begin{align}
    \hat{K}_{\mathrm{MF}}(Q) =&\; \int d\bm{k}\, \xi(\bm{k};Q) \hat{c}_{\bm{k}}^{\dagger} \hat{c}_{\bm{k}}\nonumber \\
    &+ \frac{1}{2} \int d\bm{k}\, \Big(\Delta(\bm{k};Q) \hat{c}_{\bm{k}}^{\dagger} \hat{c}_{-\bm{k}}^{\dagger} + \mathrm{H.c.}\Big) + E_{\mathrm{pair}}(Q),
    \label{Kmf}
\end{align}
where $E_{\mathrm{pair}}(Q)$ is a $c$-number energy produced by the mean-field decoupling, and the electronic operators at $\bm{k}$ and $-\bm{k}$ denote paired time-reversed states. The first term involves the band energy measured relative to the Fermi level,
\begin{align}
    \xi(\bm{k};Q) = \varepsilon_{n_0}(\bm{k};Q) - \mu(Q),
    \label{xi}
\end{align}
and the second term features the gap function $\Delta(\bm{k};Q)$ that is determined self-consistently from the gap equation
\begin{align}
    \Delta(\bm{k};Q) = - \int \frac{d\bm{p}}{(2\pi)^d} \mathcal{V}(\bm{k},\bm{p};Q) \chi(\bm{p};Q,T_c(Q)) \Delta(\bm{p};Q),
    \label{gapeqn}
\end{align}
where the pair susceptibility is
\begin{align}
    \chi(\bm{p};Q,T) \equiv \frac{1}{2\abs{\xi(\bm{p};Q)}} \tanh(\frac{\abs{\xi(\bm{p};Q)}}{2 k_B T}).
    \label{Lfunction1}
\end{align}
This gap equation can be used to determine the critical temperature $T = T_c(Q)$ at which the superconducting phase transition occurs; it should be read as follows. For each fixed value of $Q$, one obtains a distinct BCS gap equation, and solving that equation determines whether a ferroelectric background labelled by $Q$ supports superconductivity. If it does, then this equation determines the critical temperature and the symmetry of the incipient gap function associated with that background. Taking $Q = Q_0$ to be the physical order parameter (\ref{Q0orderparameter}) for a given material will determine whether or not that material supports superconductivity. 

The gap equation (\ref{gapeqn}) can be organized into pairing channels \cite{Sigrist1991,Frigeri2004}. From the point group of the crystal's lattice $\Gamma$, we find a symmetry-adapted basis $f_{\gamma}(\bm{k})$ with $\gamma$ labelling the irreducible representations of that point group; under an isotropic reduction of this group, the index $\gamma$ labels the canonical pairing channels ($s$-wave, $p$-wave, and so on). The gap function is expanded as a linear combination of these symmetry-adapted functions, and, in general, many such pairing channels exist, and can even become entangled. But we are going to consider the simpler scenario where there is only one pairing channel $\gamma$ that dominates the Cooper instability, in which case the gap function can be written
\begin{align}
    \Delta(\bm{k};Q) = \eta_{\gamma}(Q) f_{\gamma}(\bm{k}), \qquad \eta_{\gamma}(Q) \in \mathbb{C},
\end{align}
where we choose the normalization 
\begin{align}
    \int \frac{d\bm{k}}{(2\pi)^d} \abs{f_{\gamma}(\bm{k})}^2 = 1.
    \label{normalization}
\end{align}
Then the gap equation (\ref{gapeqn}) becomes
\begin{align}
    \lambda_{\gamma}(Q,T_c(Q)) = 1,
    \label{Gaplambda}
\end{align}
where the \textit{channel eigenvalue} $\lambda_{\gamma}(Q,T)$ is defined by
\begin{align}
    \lambda_{\gamma}(Q,T)f_{\gamma}(\bm{k}) = - \int \frac{d\bm{p}}{(2\pi)^{d}} \mathcal{V}(\bm{k},\bm{p};Q) \chi(\bm{p};Q,T) f_{\gamma}(\bm{p}).
    \label{channeleigenvalue}
\end{align}
Superconductivity in the channel $\gamma$ exists precisely when there exists a nonzero critical temperature $T = T_c(Q)$ satisfying the constraint (\ref{Gaplambda}). The physical superconducting phase transition is then obtained by evaluating this condition at the physical order parameter (\ref{Q0orderparameter}).

\section{Practical considerations }\label{Sec:Practicalconsiderations}

An advantage of this ferroelectric BCS model is that we can present simple analytic conditions for when ferroelectricity and superconductivity can coexist, and, in particular, when it may be enhanced in the ferroelectric phase \cite{Edge2015,Saha2025}. As discussed in Sec.~\ref{Sec:Pairing}, each value of the real parameter $Q$ defines a different uniform polar background, and therefore a different linearized gap equation. The corresponding critical temperature will be denoted by $T_c(Q)$. For a material whose ferroelectric order parameter is given by Eq. (\ref{Q0orderparameter}), we say that superconductivity is locally enhanced by an increase in the polar distortion when 
\begin{align}
    \eval{\frac{\mathrm{d}T_c}{\mathrm{d}Q}}_{Q = Q_0} > 0.
    \label{enhancesuppressTc}
\end{align}
Here and below we work in one of the two inversion-related ferroelectric domains, so that $Q$ may be taken to measure the amplitude of this polar distortion. The physical scalar quantities entering the Cooper instability problem are the same in domains with $\pm Q_0$. 

The critical temperature is determined by the gap equation (\ref{Gaplambda}). For the attractive channel considered here, the channel eigenvalue decreases at the transition, $\partial_T \lambda_{\gamma}(Q,T_c(Q)) < 0$, because the pair susceptibility (\ref{Lfunction1}) decreases with temperature. Thus, the total derivative in the enhancement condition (\ref{enhancesuppressTc}) gives
\begin{align}
    \eval{\frac{\partial \lambda_{\gamma}(Q,T)}{\partial Q}}_{(Q,T) = (Q_0,T_c(Q_0))} > 0.
    \label{enhancesuppressTcmodified}
\end{align}
This also provides us with an interpretation of the eigenvalue criterion (\ref{Gaplambda}). At a fixed temperature, the normal state becomes unstable when the largest eigenvalue of the gap equation reaches unity. If the polar distortion increases this eigenvalue at the critical temperature for the superconducting phase transition, the equality (\ref{Gaplambda}) is reached at a higher temperature and $T_c$ is raised. If it decreases the channel eigenvalue, the transition is suppressed. These results are general within the single-channel reduction of Sec.~\ref{Sec:Pairing}; to obtain some simple conditions, we now specialize to a minimal model.

\subsection{$s$-wave pairing in a parabolic band}

We focus on a spin-degenerate, isotropic parabolic band in $d = 2$ dimensions where the pairing channel is of $s$-wave symmetry. In terms of an appropriate $s$-wave basis function $f_{s}(\bm{k})$ that will be determined below, we can write the gap equation (\ref{Gaplambda}) as
\begin{align}
    1 = - \int \frac{d\bm{k}}{(2\pi)^{d}}\frac{d\bm{p}}{(2\pi)^{d}}\, f_s^*(\bm{k}) \mathcal{V}(\bm{k},\bm{p};Q) \chi(\bm{p};Q,T_c(Q)) f_s(\bm{p}).
    \label{s_gapeqn}
\end{align}
We consider energies in a small region around the Fermi surface, defined by a cutoff $\Lambda(Q)$ that represents the energies relevant to pairing; this will be determined by the Debye frequency in the ferroelectric phase. We assume a parabolic band dispersion in this region,
\begin{align}
    \xi(\bm{k};Q) = \frac{\hbar^2}{2m_*(Q)}\big(k^2 - k_F^2\big),
    \label{parabolicband}
\end{align} 
where $m_*(Q)$ is the effective mass of the retained band $n_0$ in our ferroelectric BCS model. For a fixed carrier density $n_{2d}$, this gives a circular Fermi surface of radius $k_F = \sqrt{4\pi n_{2d}/g}$ with $g = 2$ due to the spin degeneracy. The Fermi velocity and the single-spin density of states are then given by
\begin{align}
    v_F(Q) = \frac{\hbar k_F}{m_*(Q)},\qquad N_{F}(Q) = \frac{m_*(Q)}{2 \pi \hbar^2},
\end{align}
so the entire dependence of the retained band $n_0$ on the parameter $Q$ that defines the polar distortion is encoded in the effective mass of this band. The simplest $s$-wave basis function for this parabolic band model is constant inside the energy region determined by $\Lambda(Q)$ and zero outside of it, which can be written as
\begin{align}
    f_{s}(\bm{k};Q) = \frac{1}{\sqrt{2\Lambda(Q)N_F(Q)}}\Theta(\Lambda(Q) - \abs{\xi(\bm{k};Q)}),
    \label{fsbasis}
\end{align}
where the prefactor is fixed by the normalization (\ref{normalization}). Since we focus on $s$-wave pairing, and since the pair susceptibility (\ref{Lfunction1}) is strongly peaked at the Fermi energy, we restrict the momenta $(\bm{k},\bm{p})$ in the interaction kernel to the circular Fermi surface, parametrized by a pair of angles $(\theta_{\bm{k}},\theta_{\bm{p}})$. Thus, the interaction kernel will be evaluated between two points $(\theta_{\bm{k}},\theta_{\bm{p}})$ on the Fermi surface, and the gap equation (\ref{s_gapeqn}) with the basis function (\ref{fsbasis}) simplifies to
\begin{align}
    1 = g_s(Q) \int_{-\Lambda(Q)}^{\Lambda(Q)} d\xi\,\chi(\xi;Q,T_c(Q)),
    \label{swave_simplified_gapeqn}
\end{align}
involving the dimensionless $s$-wave attraction strength
\begin{align}
    g_s(Q) = -N_F(Q) \int_0^{2\pi} \frac{d\theta_{\bm{k}}}{2\pi} \int_0^{2\pi} \frac{d\theta_{\bm{p}}}{2\pi} \mathcal{V}(\theta_{\bm{k}},\theta_{\bm{p}}; Q).
    \label{swaveattraction}
\end{align}
When $k_B T_c(Q) \ll \Lambda(Q)$ we can take the asymptotic limit of the energy integral,
\begin{align}
    \int_{-\Lambda(Q)}^{\Lambda(Q)} d\xi\, \chi(\xi;Q,T_c(Q)) \approx \ln\left({2 e^{\gamma_E} \over \pi} {\Lambda(Q) \over k_B T_c(Q) }\right),
    \label{BCSasymptotic}
\end{align}
where $\gamma_{E}$ is the Euler-Mascheroni constant. We are then left with the task of evaluating the dimensionless $s$-wave attraction strength (\ref{swaveattraction}). This consists of two contributions, coming from the Coulomb and phonon-mediated parts of the interaction kernel (\ref{BCSinteraction}), which we consider separately below. 

We begin with the screened Coulomb interaction. Our ferroelectric BCS model is based on the assumption that there is only a single band $n_0$ that dominates the Cooper instability, but the remaining bands $n \neq n_0$ can still influence the system's properties. We account for this by means of a background dielectric constant $\varepsilon_{\infty}(Q)$, which depends on the parameter $Q$ because the polar distortion it identifies will generally modify the dynamics of these ``background" electrons. Since these electrons are assumed to participate in screening instead of Cooper pairing, the natural way to include this screening is through the Thomas-Fermi approximation \cite{Stern1967}, taking the screened Coulomb interaction kernel to be
 \begin{align}
     \mathcal{V}^{(c)}(\bm{k},\bm{p};Q) \approx \frac{2\pi e^2}{\varepsilon_{\infty}(Q)(q + q_{TF}(Q))},
\end{align}
involving the Thomas-Fermi wavevector
\begin{align}
    q_{TF}(Q) = \frac{2e^2}{\hbar^2}\frac{m_*(Q)}{\varepsilon_{\infty}(Q)}.
\end{align}
Since this interaction depends only on the relative angle $\phi = \theta_{\bm{p}} - \theta_{\bm{k}}$ through $q = 2k_F \sin(\phi/2)$, we can evaluate the Fermi-surface integral in the Coulomb contribution to the attraction strength (\ref{swaveattraction}), which is found to be
\begin{align}
    g_{s}^{(c)}(Q) = - \frac{e^2}{\pi\hbar^2 k_F} \frac{m_*(Q)}{\varepsilon_{\infty}(Q)}F_{c}\big(x_{TF}(Q)\big),
    \label{Coulombgs}
\end{align}
where we have defined
\begin{align}
    F_c(x) \equiv \int_0^{\pi/2} \frac{d\varphi}{\sin\varphi + x},\qquad x_{TF}(Q) = \frac{q_{TF}(Q)}{2k_F},
\end{align}
and we have set $\varphi = \phi/2$.

Next we consider the second term in the BCS interaction kernel (\ref{BCSinteraction}), which is the attractive contribution coming from the phonon-mediated interaction. Assuming a time-reversal symmetric band, this is given by
\begin{align}
    \mathcal{V}^{(ph)}(\bm{k},\bm{p};Q) = - \sum_{\lambda} \frac{1}{\Omega_{\lambda}(\bm{q};Q)^2} \abs{G_{\lambda}(\bm{k},\bm{q};Q)}^2.
\end{align}
This involves a sum over all $2N_{\mathrm{ion}}$ branches, where $N_{\mathrm{ion}}$ is the number of ions in the unit cell. Of those, there are $2$ acoustic branches and $2N_{\mathrm{ion}} - 2$ optical branches, where the branch $\lambda = \lambda_0$ that undergoes spontaneous symmetry breaking in the ferroelectric phase transition is assumed to be an optical one. We assume that the polar branch $\lambda_0$ provides the dominant contribution to this phonon-mediated interaction; this can be expected near a ferroelectric instability provided that the polar branch remains appreciably coupled to the retained electronic band. Thus, in our minimal model we take
\begin{align}
    \mathcal{V}^{(ph)}(\bm{k},\bm{p};Q) \approx - \frac{1}{\Omega_{\lambda_0}(\bm{q};Q)^2} \abs{G_{\lambda_0}(\bm{k},\bm{q};Q)}^2.
    \label{phononVsimplified}
\end{align}
We approximate the phonon dispersion for this branch in the paraelectric phase by an isotropic model with stiffness $\kappa_0$ around the zone center ($\bm{q} = \bm{0}$), in which case
\begin{align}
    \omega_{\lambda_0}(\bm{q})^2 \approx \omega_{\lambda_0}(\bm{0})^2 + \kappa_0^2 q^2.
    \label{parelectricdispersionphonon}
\end{align}
Assuming the quartic coefficient in the renormalized frequency (\ref{shiftedfrequencies}) is effectively local in momentum space, the ferroelectric dispersion relation is then given by
\begin{align}
    \Omega_{\lambda_0}(\bm{q};Q)^2 = \Omega_0(Q)^2 + \kappa_0^2 q^2,
    \label{qdispersion}
\end{align}
where the leading term is
\begin{align}
    \Omega_{0}(Q)^2 = \omega_{\lambda_0}(\bm{0})^2 + \frac{1}{2} \Gamma_4 Q^2,
    \label{Omega0Q}
\end{align}
and $\Gamma_4 \equiv \Gamma_{4}^{\lambda_0\lambda_0\lambda_0\lambda_0}(\bm{0},\bm{0},\bm{0},\bm{0}) > 0$ is required to stabilize the double-well potential for the structural phase transition. Then the interaction kernel (\ref{phononVsimplified}) becomes
\begin{align}
    \mathcal{V}^{(ph)}(\bm{k},\bm{p};Q) \approx - \frac{1}{\Omega_0(Q)^2 + \kappa_0^2 q^2} \abs{G_{\lambda_0}(\bm{k},\bm{q};Q)}^2.
    \label{phononVsimplified2}
\end{align}

We simplify the electron-phonon vertex factors (\ref{electronphononvertex}), where only the Coulomb-gradient coupling in Eq. (\ref{Gtensorcontributions}) is retained, by writing them in a reciprocal-lattice expansion of the form
\begin{align}
    G_{\lambda_0}(\bm{k},\bm{q};Q) =&\; \frac{ie}{\mathcal{N}_{\mathrm{uc}}} \sum_{\bm{G}} (\bm{q} + \bm{G})\cdot\bm{\mathcal{Z}}_{\lambda_0}(\bm{q}+\bm{G};Q)\nonumber \\
    &\times U(\bm{q}+\bm{G};Q)\mathcal{M}(\bm{k},\bm{q},\bm{G};Q),
    \label{vertex2}
\end{align}
where we have introduced the effective charge vector
\begin{align}
    \bm{\mathcal{Z}}_{\lambda_0}(\bm{q};Q) = \sum_{I} \frac{q_I}{\sqrt{M_I}} e^{-i\bm{q}\cdot(\bm{d}_I+\bm{u}_I(Q))} \bm{e}_{I\lambda_0}(\bm{q})
\end{align}
for the polar branch, and the electronic form factor
\begin{align}
    \mathcal{M}(\bm{k},\bm{q},\bm{G};Q) = \frac{1}{\Omega_{\mathrm{uc}}} \int_{\Omega} d\bm{x}\, e^{i\bm{G}\cdot\bm{x}} u_{n_0,\bm{k}+\bm{q}}^{\dagger}(\bm{x};Q) u_{n_0,\bm{k}}(\bm{x};Q).
\end{align}
The term with $\bm{G} = \bm{0}$ is the spatially smooth macroscopic component of the electron-phonon coupling, while the $\bm{G} \neq \bm{0}$ components correspond to local-field corrections that sample the internal structure of the unit cell. We retain only the $\bm{G} = \bm{0}$ term, assuming the electrons predominantly couple to the fluctuations of the \textit{macroscopic} polarization field of the ions. We also assume that the metal is sufficiently dilute so that the momentum transfer obeys $qa \leq 2k_F a \ll 1$, allowing this term to be expanded around the $\bm{q} = \bm{0}$ zone center. Then, in the Thomas-Fermi approximation, the electron-phonon vertex factor (\ref{vertex2}) becomes
\begin{align}
    G_{\lambda_0}(\bm{k},\bm{q};Q) \approx \frac{G_{0}}{\varepsilon_{\infty}(Q)} \frac{q}{q + q_{TF}(Q)}\big(\hat{\bm{q}}\cdot\hat{\bm{n}}\big)\mathcal{M}(\bm{k},\bm{q};Q),
\end{align}
where $\hat{\bm{n}}$ is the polar axis that is spontaneously selected in the broken-symmetry phase, and we have introduced the long-wavelength coupling 
\begin{align}
    G_0 = \frac{2\pi e}{\mathcal{N}_{\mathrm{uc}}} \sum_{I} \frac{q_I}{\sqrt{M_I}} \hat{\bm{n}}\cdot\bm{e}_{I\lambda_0}(\bm{0}),
    \label{G0}
\end{align}
and also the $\bm{G} = \bm{0}$ electronic form factor
\begin{align}
    \mathcal{M}(\bm{k},\bm{q};Q) = \frac{1}{\Omega_{\mathrm{uc}}} \int_{\Omega} d\bm{x}\, u_{n_0,\bm{k}+\bm{q}}^{\dagger}(\bm{x};Q) u_{n_0,\bm{k}}(\bm{x};Q),
    \label{Mformfactor}
\end{align}
Under these approximations, the phonon-mediated interaction kernel (\ref{phononVsimplified}) becomes
\begin{align}
    \mathcal{V}^{(ph)}(\bm{k},\bm{p};Q) \approx&\; - \frac{\abs{G_0}^2}{\varepsilon_{\infty}(Q)^2} \big(\hat{\bm{q}}\cdot\hat{\bm{n}}\big)^2 \abs{\mathcal{M}(\bm{k},\bm{q};Q)}^2\nonumber \\
    &\times \frac{1}{\Omega_{0}(Q)^2 + \kappa_0^2 q^2}\left(\frac{q}{q + q_{TF}(Q)}\right)^2.
\end{align}
The electronic form factor (\ref{Mformfactor}) is normalized in the sense that $\mathcal{M}(\bm{k},\bm{0};Q) = 1$. After factoring out the trivial spin degeneracy, we assume that the retained band is isolated and nondegenerate, so that the cell-periodic Bloch function can be chosen smoothly in a neighborhood of the Fermi surface. Then a Taylor expansion around $\bm{q} = \bm{0}$ leads to
\begin{align}
    \abs{\mathcal{M}(\bm{k},\bm{q};Q)}^2 = 1 - g_{ij}(\bm{k};Q) q^i q^j + O(q^3 a^3),
\end{align}
where $g_{ij}(\bm{k};Q)$ is the Abelian quantum metric for the retained band. This second term is of order $O(q^2 a^2)$ and can be dropped when the cell-periodic Bloch functions vary over momentum scales of order $a^{-1}$ and $qa \ll 1$, where the second inequality is satisfied in our dilute-metal approximation. However, it should be emphasized that we are not dropping all of the dependence on $\bm{k}$, since the unit vector $\hat{\bm{q}}$ still depends on the angles $(\theta_{\bm{k}},\theta_{\bm{p}})$. We take the unit vector $\hat{\bm{n}}$ to lie in the conducting plane, in which case rotational symmetry yields the angular average $\langle(\hat{\bm{q}}\cdot\hat{\bm{n}})^2\rangle = 1/2$ in $d = 2$ dimensions. Then the phonon contribution to the dimensionless $s$-wave attraction strength is found to be
\begin{align}
     g_{s}^{(ph)}(Q) = \frac{\abs{G_0}^2}{2\pi\hbar^2} \frac{m_*(Q)}{\varepsilon_{\infty}(Q)^2 \Omega_0(Q)^2} F_{ph}\big(x_{TF}(Q), \eta(Q)\big),
    \label{phonongs}
\end{align}
where $\eta(Q) \equiv 2 \kappa_0 k_F / \Omega_0(Q)$ and we have defined
\begin{align}
    F_{ph}(x,\eta) \equiv \frac{1}{\pi} \int_0^{\pi/2} d\varphi \left(\frac{\sin\varphi}{x + \sin\varphi}\right)^2 \frac{1}{1 + \eta^2 \sin^2\varphi}.
    \label{Fph}
\end{align}

Combining the Coulomb contribution (\ref{Coulombgs}) and this phonon-mediated contribution, the dimensionless $s$-wave attraction strength (\ref{swaveattraction}) in this minimal model is
\begin{align}
    g_{s}(Q) =&\; \frac{\abs{G_0}^2}{2\pi\hbar^2} \frac{m_*(Q)}{\varepsilon_{\infty}(Q)^2 \Omega_0(Q)^2} F_{ph}\big(x_{TF}(Q), \eta(Q)\big)\nonumber \\
    &- \frac{e^2}{\pi\hbar^2 k_F} \frac{m_*(Q)}{\varepsilon_{\infty}(Q)}F_{c}\big(x_{TF}(Q)\big).
    \label{FSschannel}
\end{align}
For $g_s(Q) > 0$, the asymptotic formula (\ref{BCSasymptotic}) gives an estimate for the critical temperature of the superconducting phase transition,
\begin{align}
    k_B T_c(Q) \simeq \frac{2 e^{\gamma_{E}}}{\pi} \Lambda(Q) \exp(-\frac{1}{g_s(Q)}).
    \label{Tcestimate}
\end{align}
The critical temperature for a given material can be estimated by evaluating this at the physical order parameter (\ref{Q0orderparameter}), and the local enhancement criterion (\ref{enhancesuppressTcmodified}) becomes
\begin{align}
    \eval{\frac{d g_s(Q)}{d Q} + g_s(Q)^2 \frac{d \ln\Lambda(Q)}{d Q}}_{Q = Q_0} > 0.
    \label{Enhancmentcriterion}
\end{align}
We take the energy cutoff $\Lambda(Q)$ to be the Debye energy $\hbar\Omega_D(Q)$ in the ferroelectric phase, where the Debye frequency $\Omega_D(Q)$ for the polar branch is estimated by matching the total number of modes in the usual Debye construction; the Debye momentum for the isotropic phonon dispersion (\ref{parelectricdispersionphonon}) is $q_D = \sqrt{4\pi/\Omega_{\mathrm{uc}}}$, and then it is straightforward to show that
\begin{align}
    \Omega_D(Q) = \sqrt{\omega_{D}^2 + \frac{1}{2} \Gamma_4 Q^2},
    \label{DebyeOmega}
\end{align}
where $\omega_{D}$ is the Debye frequency of the phonon branch $\lambda_0$ in the paraelectric phase before it becomes polar.

\subsection{Sample calculation}

\begin{figure}[h]
    \centering
    \includegraphics[width=\linewidth]{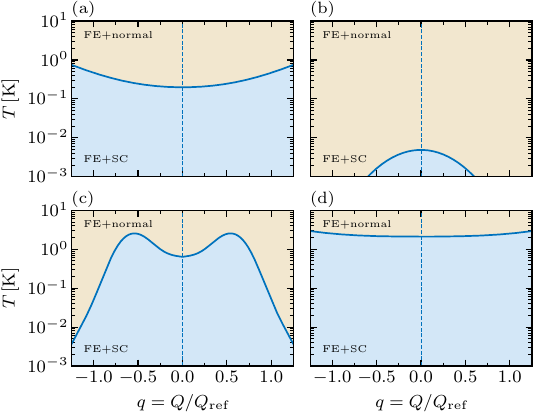}
    \caption{Phase diagram for (a) enhanced pairing, (b) suppressed pairing, (c) peaked enhancement, and (d) high--$T_c$ plateau. We use the parameters in Appendix \ref{Appendix:Parameters}.}
    \label{Fig:Phasediagram}
\end{figure}

\begin{figure}[h]
    \centering
    \includegraphics[width=\linewidth]{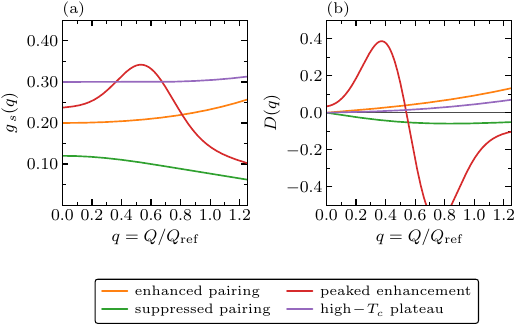}
    \caption{Representative evaluations of the ferroelectric BCS model as functions of the dimensionless polar coordinate $q$.
    (a) The $s$-wave attraction strength $g_s(q)$.
    (b) The local enhancement diagnostic
    $D(q) \equiv \frac{dg_s}{dq}+g_s(q)^2\frac{d\ln\Lambda}{dq}$ in Eq.~(\ref{Enhancmentcriterion});
    positive values correspond to local enhancement of $T_c(q)$ with increasing polarity.}
    \label{Fig:Channelplots}
\end{figure}

To illustrate this ferroelectric BCS model, we perform some simple calculations of the attraction strength (\ref{FSschannel}), the critical temperature (\ref{Tcestimate}), and the enhancement criterion (\ref{Enhancmentcriterion}) for a few different archetypal scenarios. We do so by introduction of appropriate parametrizations for the effective mass and dielectric function, the details of which can be found in Appendix \ref{Appendix:Parameters}. We choose numerical values for these parametrizations that produce pairing enhancement motivated by experiments on SrTiO$_3$-based systems \cite{Ahadi2019,Hameed2022}, together with an illustrative pairing-suppression case, along with the possibility of a high-$T_c$ plateau for interfacial systems \cite{Zhang2026,Dong2026}, and a peaked enhancement qualitatively motivated by sliding ferroelectrics \cite{Jindal2023}. These parameters are motivated by the materials discussed in the Introduction, and are also discussed in Appendix \ref{Appendix:Parameters}. The phase diagrams for each of these model archetypes are given in Figure \ref{Fig:Phasediagram}, and plots of the above quantities are shown in Figure \ref{Fig:Channelplots}.

Since the normalization of the ferroelectric order parameter depends on the convention used for the polar normal mode, the figures are plotted in terms of the dimensionless coordinate $q = Q / Q_{\mathrm{ref}}$ with $Q_{\mathrm{ref}}$ being a fixed reference amplitude. For conventional ferroelectrics with a polar distortion that is intrinsic to the lattice, our two-parameter minimal model consisting of a parametrized effective mass $m_*(q)$ and dielectric function $\varepsilon_{\infty}(q)$ with a constant polar electron-phonon coupling (\ref{G0}) is sufficient. However, for sliding ferroelectrics where the polar distortion arises from a sliding displacement that can strongly modify the layer character of the electronic states \cite{Jindal2023,Chaudhary2024}, the coupling between the polar branch and the electronic band should depend on the ferroelectric background, so we replace (\ref{G0}) with an effective background-dependent coupling $G_{\mathrm{eff}}(q)$ (see Appendix \ref{Appendix:Parameters}). 

We calculate the $s$-wave attraction strength (\ref{FSschannel}) for the archetypal scenarios above, shown in Figure \ref{Fig:Channelplots}. As the strength of the polar distortion described by $Q$ is varied, the screened Coulomb and electron-phonon contributions to the attraction strength change, and superconductivity is favoured when the phonon-mediated attraction grows relative to the screened Coulomb repulsion. This may occur through an enhanced density of states, stronger dielectric screening, softening of the relevant polar branch, or an increased polar electron-phonon matrix element. Conversely, superconductivity is suppressed when the same distortion hardens the phonon branch, lowers the density of states, weakens the effective screening, or increases the repulsive part of the interaction. 

Estimates for the corresponding transition temperatures obtained from Eq. (\ref{Tcestimate}) are shown for the same parameter choices. The trends of the curves for $T_c(q)$ are amplified compared to those of the attraction strength $g_s(q)$, because the weak-coupling critical temperature depends exponentially on the inverse pairing strength. This means a mild increase in $g_s(q)$ for the enhancement curve, motivated by conventional ferroelectrics, gives a visible increase in $T_c(q)$, while the maximum in the peaked enhancement curve, which is motivated by a sliding ferroelectric, becomes sharper in this figure. This is why a small polar correction to the microscopic interaction can still produce a significant change in the superconducting critical temperature.

The local version of this balance is more visible in the enhancement criterion of Eq. (\ref{Enhancmentcriterion}). The zeros of this function mark extrema of $T_c(q)$ in the reduced model, while a positive value means that the polar distortion locally enhances superconductivity, and a negative value means that it locally suppresses superconductivity. The first term on the left-hand side of Eq. (\ref{Enhancmentcriterion}) measures the change in the effective pairing strength, while the second accounts for the change in the pairing cutoff. This distinction is important because ferroelectricity can increase the strength of the interaction kernel while simultaneously reducing the energy window for pairing.


\section{Discussion}\label{Sec:Discussion}

We have introduced a theoretical description of crystalline metals that can host simultaneous ferroelectric and
superconducting phases \cite{Rischau2017,Hameed2022,Jindal2023,Zhang2026,Dong2026}. The electronic and ionic degrees of freedom are both described at the microscopic level, where the former are encoded in terms of a second-quantized $2$-spinor electron field operator, while the latter are encoded in first-quantized position and momentum operators for each ion individually. The ionic charge density is used to define a microscopic polarization field for the ions that couples to the electron field operator through the screened Coulomb interactions between the electrons and ions, and also spin-orbit coupling. In this sense, the formalism of Sec.~\ref{Sec:Fieldtheory} provides a fundamental description of the coupled electron-ion dynamics in terms of which the macroscopic effects of ferroelectricity and superconductivity emerge, instead of introducing the ferroelectric order as a macroscopic order parameter that is coupled phenomenologically to the superconducting gap \cite{Ginzburg1950,Sigrist1991}. 

Ferroelectricity occurs when inversion symmetry is spontaneously broken by a rearrangement of the equilibrium lattice configuration across the ferroelectric phase transition, resulting in a \textit{macroscopic} spontaneous polarization. Our description of ferroelectricity in Sec.~\ref{Sec:Ferroelectrics} is very general, applicable to both conventional ferroelectrics where the polar distortion is intrinsic to the lattice, and sliding ferroelectrics where the macroscopic polarization arises from a relative displacement between layers \cite{Rischau2017,Ahadi2019,Hameed2022,Fei2018,DeLaBarrera2021,Jindal2023}. We analyzed the spontaneous breaking of inversion symmetry from a centrosymmetric paraelectric phase by making the approximation that the equilibrium configuration in the ferroelectric phase involves a uniform spontaneous polarization, which is the common scenario in ferroelectric materials at sufficiently low temperatures. Quantized normal-mode oscillations about this ferroelectric configuration are phonons, and by integrating out these phonons we obtained an effective electronic Hamiltonian that features a $2$-body interaction term accounting for both the screened Coulomb interactions and the phonon-mediated interactions between the electrons. 

When the kernel for this effective interaction term attains a negative eigenvalue, implying that the attractive phonon-mediated interactions dominate over the repulsive Coulomb interactions, the corresponding ``eigenchannel" becomes a pairing channel that undergoes the phonon-mediated Cooper instability and becomes superconducting. Since the effective Hamiltonian we derived in Sec.~\ref{Sec:Pairing} involves the full spin structure of the electronic sector along with generic spin-orbit coupling, our treatment is applicable to any kind of phonon-mediated instability, not only the spin-independent $s$-wave instability that underlies conventional BCS theory. Indeed, our effective Hamiltonian involves a general spin-dependent, possibly multiband interaction kernel whose attractive eigenchannel(s) may represent singlet, triplet, mixed-parity, or even spin-orbit-entangled pairing states. In this sense our formalism lends itself naturally to analyses based on \textit{ab initio} numerical methods such as density functional theory, and may be used to search for new materials that support ferroelectric superconductivity \cite{Baroni2001,Giustino2017,HickoxYoung2023}. 

To study the superconducting phase transition in the ferroelectric phase, we reduced our general multiband Hamiltonian to a single-band, spin-independent ferroelectric BCS model, where the polar distortion enters as a parameter describing a family of BCS models that depends continuously on the ferroelectric background through this parameter. This produces a one-parameter family of gap equations for the superconducting order parameter that determines the critical temperature of the phase transition in each such background, and the question of whether ferroelectricity enhances superconductivity becomes a question about whether an increase in the polar distortion increases the pairing strength. In the weak-coupling limit, this dependence is amplified exponentially in the critical temperature, implying that even a modest polar modification of the BCS interaction kernel can produce a relatively substantial change in the critical temperature. Our focus concerned the spin-independent pairing mechanisms of conventional BCS theory, but this construction can be relaxed: A more general spin structure with spin-orbit coupling can be included by an appropriately modified reduction of the interaction kernel (\ref{Vintkernel}), with which one may derive BCS-type models for spin-dependent, parity-odd pairing \cite{Gorkov2001,Frigeri2004,Kanasugi2018,Kozii2019}. This is a topic for future work.

The calculations in Sec.~\ref{Sec:Practicalconsiderations} demonstrate how the problem of enhancement or suppression of superconductivity can be resolved in our ferroelectric BCS model. The polar distortion simultaneously modifies the density of states, dielectric screening, phonon spectrum, pairing window, and couplings, and a ferroelectric enhancement of superconductivity occurs only when these modifications collectively increase the attractive pairing strength (\ref{FSschannel}) enough to overcome any reduction of the energy window in which Cooper pairing occurs \cite{Edge2015,Dunnett2018,VanDerMarel2019,Gastiasoro2020,Saha2025}. This is why the local diagnostic (\ref{Enhancmentcriterion}) is useful: It separates the change in the pairing ``strength" from the change in the pairing window. The conventional and sliding cases then differ primarily in which of the microscopic ingredients that comprise the attraction strength (\ref{FSschannel}) are most sensitive to the ferroelectric background. For conventional ferroelectrics, such as strained $\mathrm{SrTiO}_3$ with chemical substitution, the dominant effects are naturally captured by an effective mass, a background dielectric response, and the polar-phonon frequency \cite{Rischau2017,Ahadi2019,Russell2019,Hameed2022}. However, for sliding ferroelectrics, including bilayer $\mathrm{WTe}_2$ and $T_d$-$\mathrm{MoTe}_2$ with appropriate stacking, the relative layer displacement can also substantially modify the electronic wavefunctions through virtual charge transfer between layers, and thereby modify the coupling of the polar mode to the electronic sector \cite{Fei2018,DeLaBarrera2021,Yang2018,Liu2019,Jindal2023,Chaudhary2024,Biswas2026,Annenkov2025}. 

The formalism introduced here provides a quantitative foundation to a simple yet important observation: The ferroelectric order may enhance or suppress superconductivity depending on how a change in the background lattice structure redistributes the balance between Coulomb repulsion, phonon-mediated attraction, and the available pairing phase space. And by constructing this foundation at the microscopic level, we provide a route for connecting the symmetry-breaking physics of ferroelectrics to the pairing problem of superconductivity in both conventional and sliding ferroelectric metals.


\begin{acknowledgments}
    This work was supported by the Natural Sciences and Engineering Research Council of Canada (NSERC). We thank Ivar Martin and Gaurav Chaudhary for the interesting and useful discussions. 
\end{acknowledgments}


\appendix


\section{Effective electronic Hamiltonian}\label{Sec:EffectiveHamiltonian}

We summarize the constructions leading to the effective electronic Hamiltonian (\ref{Hintkspace}). Phonons are quasiparticle excitations associated with quantized normal-mode oscillations, which in the ferroelectric phase are described by the fluctuation operators that feature in the renormalized ion Hamiltonian (\ref{ionicHfluct}) and electron-ion interaction term (\ref{Hint}). The dynamics of these phonons is described by the Hamiltonian
\begin{align}
    \hat{H}_{\mathrm{ph}}(t) \equiv \hat{H}_{\mathrm{ion}}^R(t) + \hat{H}_{\mathrm{int}}^R(t).
    \label{Hphonon}
\end{align}
We introduce bosonic creation and annihilation operators representing phononic excitations of the ferroelectric lattice configuration, writing the fluctuation operators as
\begin{align}
    \delta\hat{Q}_{\lambda}(\bm{q},t) &= \sqrt{\frac{\hbar}{2\Omega_{\lambda}(\bm{q})}} \Big(\hat{b}_{\lambda\bm{q}}(t) + \hat{b}_{\lambda,-\bm{q}}^{\dagger}(t)\Big),\nonumber \\
    \delta\hat{\Pi}_{\lambda}(\bm{q},t) &= - i \sqrt{\frac{\hbar\Omega_{\lambda}(\bm{q})}{2}} \Big(\hat{b}_{\lambda\bm{q}}(t) - \hat{b}_{\lambda,-\bm{q}}^{\dagger}(t)\Big),
    \label{normalmodeoperators}
\end{align}
where the phonon creation and annihilation operators satisfy the  canonical commutation relations. Following some straightforward operator algebra, the phonon Hamiltonian (\ref{Hphonon}) is found to be
\begin{widetext}
\begin{align}
    \hat{H}_{\mathrm{ph}}(t) = \sum_{\lambda} \int d\bm{q}\, \hbar\Omega_{\lambda}(\bm{q}) \left(\hat{b}_{\lambda\bm{q}}^{\dagger}(t)\hat{b}_{\lambda\bm{q}}(t) + \frac{1}{2}\right) - \sum_{\lambda} \int d\bm{q} d\bm{x}\, \hat{\psi}^{\dagger}(\bm{x},t) g_{\lambda}(\bm{x},\bm{q}) \hat{\psi}(\bm{x},t) \Big(\hat{b}_{\lambda\bm{q}}(t) + \hat{b}_{\lambda,-\bm{q}}^{\dagger}(t)\Big).
    \label{Hphnomix}
\end{align}
\end{widetext}
where we have defined the rescaled coupling
\begin{align}
    g_{\lambda}(\bm{x},\bm{q}) \equiv \sqrt{\frac{\hbar}{2\Omega_{\lambda}(\bm{q})}} \mathcal{G}_{\lambda}(\bm{x},\bm{q}).
    \label{gtensor}
\end{align}
This should be understood as a general spinor-valued electron-phonon coupling, because we have not yet made any assumption about the spin structure and symmetries of the electronic sector. 

To ``integrate out" the phonons, we solve the Heisenberg equation for the phonon creation and annihilation operators, and substitute their solutions back into the phonon Hamiltonian (\ref{Hphnomix}), after which one can identify an effective $2$-body electronic interaction term describing phonon-mediated interactions between the electrons. Combining it with the electronic term (\ref{electronH}), the time-nonlocal effective ``Hamiltonian" is 
\begin{widetext}
\begin{align}
    \hat{H}_{\mathrm{eff}}(t) =&\; \int d\bm{x}\, \hat{\psi}^{\dagger}(\bm{x},t) \mathcal{H}_e^R(\bm{x}) \hat{\psi}(\bm{x},t)\nonumber \\
    &+ \frac{1}{2} \sum_{\{s\}} \int_{-\infty}^{\infty} dt' \int d\bm{x}d\bm{y}d\bm{z}d\bm{w}\, \mathrm{V}_{s_1 s_2 s_3 s_4}(\bm{x},\bm{y},\bm{z},\bm{w};t-t') \hat{\psi}_{s_1}^{\dagger}(\bm{x},t) \hat{\psi}_{s_2}^{\dagger}(\bm{y},t') \hat{\psi}_{s_3}(\bm{z},t') \hat{\psi}_{s_4}(\bm{w},t),
    \label{Heff}
\end{align}
involving the Hamiltonian density (\ref{H0density}) and the effective $2$-body electronic interaction kernel 
\begin{align}
    \mathrm{V}_{s_1 s_2 s_3 s_4}(\bm{x},\bm{y},\bm{z},\bm{w};t-t') =&\; e^2 U(\bm{x} - \bm{y}) \delta_{s_1 s_4} \delta_{s_2 s_3} \delta(\bm{z} - \bm{y}) \delta(\bm{w} - \bm{x})\delta(t-t')\nonumber \\
    &- \sum_{\lambda} \int d\bm{q}\, D_{\lambda}(\bm{q};t-t')\mathcal{G}_{\lambda,s_1 s_4}(\bm{x},\bm{w};\bm{q}) \mathcal{G}_{\lambda,s_2 s_3}(\bm{y},\bm{z};-\bm{q}),
    \label{2bodyBCSkernel}
\end{align}
\end{widetext}
where the second term features the spin components of
\begin{align}
    \mathcal{G}_{\lambda}(\bm{x},\bm{z};\bm{q}) =&\; \mathcal{G}_{\lambda}^c(\bm{x},\bm{q}) \delta(\bm{x} - \bm{z})\nonumber \\
    &+ \frac{\hbar}{4m^2 c^2} \bm{\sigma} \cdot \bm{\nabla} \mathcal{G}_{\lambda}^c(\bm{x},\bm{q}) \times \bm{p}(\bm{x})\delta(\bm{x} - \bm{z}),
    \label{Gdistributions}
\end{align}
which is a distributional representation of the electron-phonon couplings (\ref{mathcalGtot}). The interaction kernel (\ref{2bodyBCSkernel}) includes memory effects through the phonon propagator
\begin{align}
    D_{\lambda}(\bm{q},t-t') = \int_{-\infty}^{\infty} \frac{d\omega}{2\pi} \frac{e^{-i\omega(t-t')}}{\Omega_{\lambda}(\bm{q})^2 - (\omega+i0^+)^2}.
    \label{phononpropagator}
\end{align}

To make the electronic band structure manifest, we use the cell-periodicity of the Hamiltonian density (\ref{H0density}) to implement Bloch's theorem, providing us with a suitable basis of single-particle electronic wavefunctions with which to expand the electron field operator, namely the Bloch energy eigenfunctions (\ref{Blocheigenfunctions}). These Bloch energy eigenfunctions (\ref{Blocheigenfunctions}) form a complete and orthonormal basis of single-particle wavefunctions, and the electron field operator admits a well-defined expansion in terms of these eigenfunctions of the form
\begin{align}
    \hat{\psi}(\bm{x},t) = \sum_{n} \int d\bm{k}\, \psi_{n\bm{k}}(\bm{x}) \hat{c}_{n\bm{k}}(t),
\end{align}
where the electronic creation and annihilation operators satisfy the canonical anticommutation relations. Implementing this expansion in Eq. (\ref{Heff}), and neglecting any retardation effects in the phonon-mediated interaction by replacing the phonon propagator (\ref{phononpropagator}) with its instantaneous approximation $D_{\lambda}(\bm{q};t-t') \to \frac{1}{\Omega_{\lambda}(\bm{q})^2} \delta(t-t')$, the momentum-space Hamiltonian (\ref{Hintkspace}) follows.

\section{Parametrizations in Sec.~\ref{Sec:Practicalconsiderations}}\label{Appendix:Parameters}

\begin{table*}[t]
\begin{ruledtabular}
\scriptsize
\setlength{\tabcolsep}{3.2pt}
\begin{tabular}{lccccccccccc}
Curve label &
$x_{TF}(0)$ &
$a_m$ &
$a_{\varepsilon}$ &
$a_{\Omega}$ &
$a_G$ &
$A_M$ &
$q_M$ &
$\sigma_M$ &
$\mu_C$ &
$g_s^{\rm ref}$ &
$\Lambda_D/k_B\,[{\rm K}]$ \\
\hline

enhanced pairing &
0.741 & 0.08 & $-0.10$ & 0.35 & 0.18 & 0 &
-- & -- & 0.070 & 0.200 & 26 \\

suppressed pairing &
0.606 & 0.06 & $-0.03$ & 0.45 & 0.08 & 0 &
-- & -- & 0.120 & 0.120 & 18 \\

peaked enhancement &
0.645 & 0.14 & $-0.08$ & 0.75 & 0.05 & 0.25 &
0.58 & 0.20 & 0.080 & 0.235 & 38 \\

high-$T_c$ plateau &
0.667 & 0.10 & $-0.09$ & 0.30 & 0.10 & 0 &
-- & -- & 0.070 & 0.300 & 52 \\

\end{tabular}
\end{ruledtabular}
\caption{
Illustrative input parameters used for the four representative
curves in Figures~\ref{Fig:Phasediagram} and
\ref{Fig:Channelplots}. The curve labels denote qualitative
behaviours of the reduced model. The material examples
discussed in the text motivate these qualitative behaviours
only; the numerical values are rough illustrative choices and
are not material-specific fits. For all curves except the
peaked-enhancement curve, $A_M=0$, and the unused Gaussian
parameters are denoted by dashes.
}
\label{tab:appendixC_input_params}
\end{table*}

We summarize the parametrizations behind Figures \ref{Fig:Phasediagram} and \ref{Fig:Channelplots} in Sec.~\ref{Sec:Practicalconsiderations}, and motivate our choice of numerical values for the associated parameters. The effective mass, dielectric function, and polar-mode frequency are taken to be functions of the dimensionless polar coordinate $q = Q / Q_{\mathrm{ref}}$, where $Q_{\mathrm{ref}}$ is a fixed reference amplitude. In the absence of any electric field that selects between the pair of backgrounds $\pm q$ that are related by an inversion transformation, these scalar quantities are assumed to be even functions, and we retain only the leading quadratic dependence,
\begin{align}
    r_m(q) \equiv \frac{m_*(q)}{m_*(0)} &= 1 + a_m q^2, \nonumber \\
    r_{\varepsilon}(q) \equiv \frac{\varepsilon_{\infty}(q)}{\varepsilon_{\infty}(0)} &= 1 + a_{\varepsilon} q^2,\nonumber \\
    r_{\Omega}(q)^2 \equiv \frac{\Omega_0(q)^2}{\Omega_0(0)^2} &= 1 + a_{\Omega} q^2.
    \label{BCSparametrization}
\end{align}
The coefficient $a_m$ encodes the change in band curvature as the background $q$ is varied, and therefore the change in the two-dimensional density of states at the Fermi surface, while the coefficient $a_{\varepsilon}$ encodes the change in the dielectric screening supplied by bands not retained in our single-band model. The coefficient $a_{\Omega}$ parametrizes the leading quadratic dependence of the squared zone-center ferroelectric fluctuation frequency $\Omega_0(Q)^2$ on the polar background, as described by Eq. (\ref{Omega0Q}). 

For conventional ferroelectrics with a polar distortion that is intrinsic to the lattice, these parametrizations with a constant polar coupling (\ref{G0}) are sufficient, but for sliding ferroelectrics where the polar distortion arises from a sliding displacement that can strongly modify the layer character of the electronic states, the coupling between the polar branch and the electronic band may depend on the ferroelectric background. Thus, we replace the scalar coupling $G_0$ with an effective background-dependent coupling $G_{\mathrm{eff}}(q)$ that is parametrized by
\begin{align}
    G_{\mathrm{eff}}(q) &= G_{\mathrm{ref}} \mathcal{G}(q),\nonumber \\
    \mathcal{G}(q) &= \big(1 + a_G q^2\big)
    \left[1+A_{M}\exp\left(-\frac{(q - q_{M})^2}{2\sigma_{M}^2}\right)\right], 
    \label{Geff}
\end{align}
for some set $\{A_{M}, q_{M}, \sigma_{M}\}$ of constant Gaussian parameters, where $q \geq 0$. The parametrization is extended to the inversion-related domain by $\mathcal{G}(-q) = \mathcal{G}(q)$. Here $G_{\mathrm{ref}}$ is a reference coupling scale whose absolute value is absorbed into the reference pairing calibration introduced below. The Gaussian factor is not necessarily a universal property of sliding ferroelectrics; rather, it is a compact parametrization of the possibility that the electron-phonon vertex is largest at an intermediate layer configuration $q = q_M$. 

For completeness, we describe how these parametrizations are inserted into the minimal ferroelectric BCS model of Sec. \ref{Sec:Practicalconsiderations}. We write the Thomas-Fermi parameter in Eq. (\ref{FSschannel}) as 
\begin{align}
    x_{TF}(q) = \frac{r_m(q)}{r_{\varepsilon}(q)} x_{TF}(0),\quad x_{TF}(0) = \frac{e^2 m_*(0)}{\hbar^2 k_F \varepsilon_{\infty}(0)},
\end{align}
and we consider the simplified scenario where $\eta(q) \approx 0$ in Eq. (\ref{phonongs}). Denote the magnitude of the Coulomb term at the reference background by $\mu_c \equiv - g^{(c)}(0) > 0$, and define the reference attraction strength by $g_{s}^{\mathrm{ref}} \equiv g_{s}(0)$. Then we find for the dimensionless $s$-wave attraction strength
\begin{align}
    g_s(q) =&\; \big(g_{s}^{\mathrm{ref}} + \mu_c\big) \frac{r_m(q) \mathcal{G}(q)^2}{r_{\varepsilon}(q)^2 r_{\Omega}(q)^2 \mathcal{G}(0)^2} \frac{F_{ph}(x_{TF}(q),0)}{F_{ph}(x_{TF}(0),0)}\nonumber \\
    &- \mu_c \frac{r_m(q)}{r_{\varepsilon}(q)} \frac{F_c(x_{TF}(q))}{F_c(x_{TF}(0))},
\end{align}
which is equivalent to Eq.~(\ref{FSschannel}), except written with respect to the reference background $q = 0$. For the cutoff in Eqs. (\ref{Tcestimate}) and 
(\ref{Enhancmentcriterion}), we use the parametrization
\begin{align}
    \Lambda(q) = \Lambda_D \sqrt{1 + a_{\Omega} q^2},
\end{align}
where $\Lambda_D \equiv \Lambda(0)$. The parametrizations above completely specify the curves in Figures \ref{Fig:Phasediagram} and \ref{Fig:Channelplots}. 

The parameter sets in Table~\ref{tab:appendixC_input_params} are chosen to generate four representative qualitative behaviours of the minimal model: Enhanced pairing, suppressed pairing, peaked enhancement, and a high-$T_c$ plateau. They are not fits to particular materials, and the numerical values should not be interpreted as measured or calculated microscopic parameters of any specific system. The material examples mentioned below serve only to motivate why these four qualitative behaviours are physically relevant.

The ``enhanced pairing" archetype is qualitatively motivated by the increase of $T_c$ reported near ferroelectricity in strained SrTiO$_3$-based systems \cite{Ahadi2019,Russell2019}. In the reduced model, this behaviour is produced by a favourable combination of increased density of states, stronger background screening, a smooth enhancement of the polar coupling, and only moderate hardening of the retained polar branch. Meanwhile, the ``suppressed pairing" archetype emphasizes that polar metallicity does not by itself guarantee enhancement of the conventional $s$-wave channel. Its qualitative motivation comes from polar metals such as few-layer WTe$_2$, where switchable polarization survives metallic screening \cite{Fei2018, DeLaBarrera2021}. In the reduced model, a harder polar branch, weaker vertex enhancement, and larger residual Coulomb penalty can instead suppress $g_s(q)$. The ``peaked enhancement" archetype is qualitatively motivated by sliding ferroelectric systems such as bilayer $T_d$-MoTe$_2$, where the polar coordinate changes the layer registry and can therefore modify the electronic states and their coupling to the polar mode \cite{Jindal2023}. The Gaussian factor in Eq.~(\ref{Geff}) is used to illustrate how a background-dependent vertex can produce a maximum at an intermediate displacement. Finally, the ``high-$T_c$ plateau" archetype is qualitatively motivated by interfacial superconducting systems such as LaAlO$_3$/KTaO$_3$, where comparatively broad superconducting regions and polar tuning have been reported \cite{Chen2021,Dong2026}. This behaviour is represented by a larger reference attraction and cutoff, a relatively small Coulomb penalty, and moderate polar dependence.

\bibliography{Bibliography}

\end{document}